\documentclass[%
reprint,
twocolumn,
superscriptaddress,
showpacs,
amsmath,amssymb,amsfonts, 
aps,
prx,
floatfix, 
notitlepage, 
]{revtex4-1}

\usepackage[dvipdfmx]{graphicx}
\usepackage{dcolumn}
\usepackage{bm}
\usepackage{braket}
\usepackage[dvipdfmx]{attachfile2}
\usepackage{braket}
\usepackage{subfigure}

\begin{abstract}
We develop the Floquet-Magnus expansion for a classical equation of motion under a periodic drive that is applicable to both isolated and open systems. 
For classical systems, known approaches based on the Floquet theorem fail due to the nonlinearity and the stochasticity of their equations of motion (EOMs) in contrast to quantum ones. 
Here, employing their master equation, we successfully extend the Floquet methodology to classical EOMs to obtain their Floquet-Magnus expansions, thereby overcoming this difficulty. 
Our method has a wide range of application from classical to quantum as long as they are described by differential equations including the Langevin equation, the Gross-Pitaevskii equation, and the time-dependent Ginzburg-Landau equation. 
By analytically evaluating the higher-order terms of the Floquet-Magnus expansion, we find that it is, at least asymptotically, convergent and well approximates the relaxation to their prethermal or non-equilibrium steady states. 
To support these analytical findings, we numerically analyze two examples: (i) the Kapitza pendulum with friction and (ii) laser-driven magnets described by the stochastic Landau-Lifshitz-Gilbert equation. 
In both cases, the effective EOMs obtained from their Floquet-Magnus expansions correctly reproduce their exact time evolution for a long time up to their non-equilibrium steady states. 
In the example of driven magnets, we demonstrate the controlled generations of a macroscopic magnetization and a spin chirality by laser and discuss possible applications to spintronics. 
\end{abstract}

\begin{document}


\newcommand{\sig}{\sigma}
\newcommand{\lam}{\lambda}
\newcommand{\wh}{\widehat}
\newcommand{\mr}{\mathrm}
\newcommand{\mf}{\mathfrak}
\newcommand{\mb}{\mathbb}
\newcommand{\tb}{\textbf}
\newcommand{\pri}{^\prime}
\newcommand{\fn}{\footnote}
\newcommand{\FOR}{ \ \mathrm{for} \ }
\newcommand{\WHEN}{ \ \mathrm{when} \ }

\newcommand{\refE}[1]{Eq.~(\ref{eq:#1})}
\newcommand{\Eref}[1]{(\ref{eq:#1})}
\newcommand{\refS}[1]{Sec.~\ref{sec:#1}}
\newcommand{\Sref}[1]{\ref{sec:#1}}
\newcommand{\refA}[1]{App.~\ref{sec:#1}}
\newcommand{\refF}[1]{Fig.~\ref{fig:#1}}
\newcommand{\Fref}[1]{\ref{fig:#1}}
\newcommand{\refT}[1]{Table~\ref{tb:#1}}
\newcommand{\refTH}[1]{Theorem~\ref{thm:#1}}
\newcommand{\refLM}[1]{Lemma~\ref{thm:#1}}
\newcommand{\refCO}[1]{Corollary~\ref{thm:#1}}
\newcommand{\COref}[1]{\ref{thm:#1}}
\newcommand{\refEX}[1]{Ex.~\ref{exa:#1}}
\newcommand{\refC}[1]{Chapter~\ref{ch:#1}}
\newcommand{\labE}[1]{\label{eq:#1}}
\newcommand{\labS}[1]{\label{sec:#1}}
\newcommand{\labF}[1]{\label{fig:#1}}
\newcommand{\labT}[1]{\label{tb:#1}}
\newcommand{\labTH}[1]{\label{thm:#1}}
\newcommand{\labEX}[1]{\label{exa:#1}}

\newcommand{\mbZ}{\mathbb{Z}}
\newcommand{\mbR}{\mathbb{R}}
\newcommand{\mbC}{\mathbb{C}}
\newcommand{\mbN}{\mathbb{N}}
\newcommand{\mbT}{\mathbb{T}}

\newcommand{\eco}{,\\}
\newcommand{\nco}{\\}
\newcommand{\nnco}{\nonumber \\}
\newcommand{\nneco}{,\nonumber \\}
\newcommand{\peqn}[1]{
\begin{align}
#1
.\end{align}
}
\newcommand{\ceqn}[1]{
\begin{align}
#1
,\end{align}
}
\newcommand{\neqn}[1]{
\begin{align}
#1
\end{align}
}
\newcommand{\twoeq}[3]{
\left\{
\begin{array}{l}
#1  \\
#2 #3
\end{array}
\right.
}
\newcommand{\eqcom}[1]{
, \ \mathrm{#1}
}

\newcommand{\tbr}[1]{
\left\{ #1 \right\}
}
\newcommand{\rbr}[1]{
\left[ #1 \right]
}
\newcommand{\br}[1]{
\left( #1 \right)
}

\newcommand{\twmat}[4]{
\left(
\begin{array}{cc}
#1 & #2 \\
#3 & #4
\end{array}
\right)
}
\newcommand{\thmat}[9]{
\left(
\begin{array}{ccc}
#1 & #2 & #3 \\
#4 & #5 & #6 \\
#7 & #8 & #9
\end{array}
\right)
}

\newcommand{\itemb}[2]{
\begin{itembox}[l]{\textbf{#1}}
#2
\end{itembox}
}
\newcommand{\bitem}[1]{
\item
\textbf{#1}\\
}

\newcommand{\Ref}[1]{$\rbr{\mr{#1}}$ \cite{#1}}
\newcommand{\bk}{\bm{k}}
\newcommand{\eps}{\epsilon}
\newcommand{\heff}{h_{\mr{eff}}}
\newcommand{\citeu}{[* Ref. *]}
\newcommand{\connl}{, \nonumber \\}
\newcommand{\change}[1]{\textcolor{red}{#1}}
\newcommand{\comm}[1]{\textcolor{blue}{(*#1*)}}
\newcommand{\twvec}[2]{\left(\begin{array}{c} #1 \\ #2 \end{array} \right)}
\newcommand{\thvec}[3]{\left(\begin{array}{c} #1 \\ #2 \\ #3 \end{array} \right)}
\newcommand{\fourvec}[2]{\left(\begin{array}{c} #1 \\ #2 #1 \end{array} \right)}

\newcommand{\wt}{\widetilde}

\newcommand{\Exp}[1]{\mathrm{e}^{#1}}
\newcommand{\mL}{\mathcal{L}}
\newcommand{\mG}{\mathcal{G}}
\newcommand{\bmr}{\bm{r}}
\newcommand{\bmphi}{\bm{\phi}}
\newcommand{\bmh}{\bm{h}}
\newcommand{\bmf}{\bm{f}}
\newcommand{\bmm}{\bm{m}}

\newcommand{\red}[1]{\textcolor{red}{#1}}




\title{Floquet engineering of classical systems}
\author{Sho Higashikawa}
\affiliation{Department of Physics, University of Tokyo, Hongo 113-8656, Japan}
\email{higashikawa@cat.phys.s.u-tokyo.ac.jp}
\author{Hiroyuki Fujita}
\affiliation{Institute for Solid State Physics, University of Tokyo, Kashiwa 277-8581, Japan}
\email{h-fujita@issp.u-tokyo.ac.jp}
\author{Masahiro Sato}
\affiliation{Department of Physics, Ibaraki University, 
Mito, Ibaraki 310-8512, Japan}
\email{masahiro.sato.phys@vc.ibaraki.ac.jp}
\date{\today}
\maketitle

\section{Introduction}
Thanks to the rapid developments in laser and ultrafast spectroscopy techniques, recent years had witnessed a remarkable progress in the studies of periodically driven quantum systems in solid-state, atomic, and molecular physics \cite{KirilyukA10, BukovM15, EckardtA17, OkaT18}, where a number of far-from-equilibrium phenomena have been investigated; 
Higgs mode in superconductors \cite{MatsunagaR13, MatsunagaR14, KatsumiK18}, dynamical localization \cite{DunlapD86, LignierH07}, and Floquet time crystals \cite{ElseD16, YaoN17, ChoiS17, ZhangJ17}, to name a few.  
Moreover, a periodic drive is found to be a new versatile tool for engineering quantum systems. 
This form of quantum engineering, usually termed \textit{Floquet engineering}, is based on the fact that the time evolution of a periodically driven quantum system is effectively described by a time-independent \textit{effective Hamiltonian} thanks to the Floquet theorem \cite{FloquetG83, ChiconeC06}, a temporal analog of the Bloch theorem. 
Floquet engineering enables us to propose and realize various non-equilibrium phenomena including the dynamic control of the superfluid-insulator transition \cite{EckardtA05, EckardtA09, ZenesiniA09}, the implementation of artificial gauge fields \cite{JakschD03, AidelsburgerM13, AidelsburgerM14, KennedyC15}, frustrated magnets \cite{StruckJ11, StruckJ13, GorgF18}, and Floquet topological insulators \cite{OkaT09, KitagawaT10b, LindnerN11, WangY13, JotzuG14, RechtsmanM13}, and the control of magnetization, spin chirality, and a spin-liquid state \cite{SatoM14, TakayoshiS14a, TakayoshiS14b, SatoM16, MentinkJ15}.

For a fast drive, the analysis of the effective Hamiltonian is simplified by a systematic expansion, which is known as the \textit{Floquet-Magnus} (FM) expansion \cite{GoldmanN14a, RahavS03, EckardtA15, KuwaharaT16, MoriT16}.
Although the FM expansion is, in general, a divergent series, its truncated series well describes the dynamics of an isolated quantum system in an intermediate time domain before heating up, which is known as the Floquet-prethermal regime \cite{KuwaharaT16, MoriT16, AbaninD15, AbaninD17}. 
It is widely believed that the divergent nature of the FM expansion is generic to non-integrable quantum many-body systems as it indicates eventual heating to a featureless infinite-temperature state due to a persistent drive \cite{DAlessioL13, DAlessioL14, PonteP15, LazaridesA14, LuitzD17, JotzuG14}. 
It has recently been shown that an isolated classical spin system under a fast periodic drive also exhibits the Floquet prethermalization similarly to quantum systems \cite{HowellO18, MoriT18}, indicating the possibility of extending the Floquet methodology to classical systems. 

While remarkable progress has been made concerning quantum systems for the last decade, periodically driven \textit{classical} systems have a long history of study, where a number of interesting phenomena have been found including dynamical localization \cite{ChirikovB81}, stochastic resonance \cite{GammaitoniL98, JungP93}, and dynamical stabilization \cite{KapitzaP51, PaulW90, CourantE52, FeynmanR64, SaitoH03, AbdullaevF03}. 
Thus, it is clearly important to extend the Floquet methodology established in quantum systems to classical ones, in particular, to develop a general framework for obtaining the FM expansion of their equation of motions (EOMs). 
In fact, such generalizations have a wide range of applications not only in purely classical systems, e.g., a Langevin system in biology and chemistry \cite{KampenV92}, but also in quantum systems in symmetry broken phases, e.g., Bose-Einstein condensates described by the Gross-Pitaevskii equation \cite{PethickC02, KurnDS13}, micromagnets described by the Landau-Lifshitz-Gilbert (LLG) equation \cite{LandauL35, GilbertT04, BrownW63}, and generic ordered phases following the time-dependent Ginzburg-Landau equation. 
Also, it is desirable to generalize it to \textit{open} classical systems for practical applications to, e.g., biology and solid-state physics. 
As emphasized in the studies on Floquet engineering of quantum systems \cite{KohnW01, GrifoniM98, HoneD09, KetzmerickR10, LangemeyerM14, IadecolaH15b, DehghaniH14, SeetharamK15, VorbergD13, VorbergD15}, the coupling with an environment is essential for preventing the system from heating up against a persistent drive. 

However, the FM expansion for classical EOMs has so far been developed only in Hamilton systems \cite{DAlessioL13, MoriT18}. 
Although the multi-scale perturbation theory has been applied to specific examples and has turned out to be successful \cite{KapitzaP51, SaitoH03, AbdullaevF03, DuttaS03, DuttaS04}, its calculation often becomes involved as is the case with the singular perturbation theory \cite{BenderC13, StrogatzS18}. 
As a result, it is difficult to analyze them in a general manner and little is known on its validity and the convergence of this perturbative expansion. 
One difficulty for the generalization is that the Floquet theorem, which lies at the heart of Floquet engineering, can be applied only to linear differential equations like Schr\"odinger equation \cite{ChiconeC06, BukovM15}, while classical systems are generally described by nonlinear EOMs. 
Another difficulty arises in open classical systems coupled to thermal reservoirs.
A thermal fluctuation appears in its EOM as a stochastic variable and breaks the exact periodicity of the EOM, which again makes the applicability of the Floquet theorem unclear. 

In this paper, we develop the FM expansion for periodically driven classical systems that is applicable to both isolated systems and open ones coupled to thermal reservoirs. 
The key idea is simple and general: using the corresponding master equations to its EOM rather than its EOM itself.
Since the master equations are linear in the probability distribution function and periodic with time, one can safely apply the Floquet theorem, thereby performing the FM expansion. 
The effective EOM is derived from the FM expansion of the master equation through the correspondence between the EOM and the master equation. 
By evaluating the higher-order terms of the FM expansion, we find that it is, at least asymptotically, convergent and well approximates the exact dynamics up to a Floquet prethermal state for an isolated system and a non-equilibrium steady state (NESS) for a driven dissipative system. 
To support these analytical findings, we numerically test its validity by two examples of open classical systems: (i) the Kapitza pendulum \cite{KapitzaP51} with friction, (ii) laser-irradiated magnets described by the stochastic LLG (sLLG) equation. 
In both cases, comparing the time evolution of the time-dependent EOMs and that of the effective ones obtained from the FM expansion, we confirm that the effective ones well approximate the time-dependent ones not only in a short time but also for a long time until their non-equilibrium steady state. 
This result is in stark contrast to isolated quantum systems where the truncated FM expansion fails to capture eventual heating to infinite temperature \cite{BukovM15, LazaridesA14, KuwaharaT16, MoriT16, AbaninD17}. 
Finally, we present an application to spintronics \cite{MaekawaS17}, where we analyze a multiferroic spin chain irradiated by a circularly polarized laser. 
We show that a synthetic Dzyaloshinskii-Moriya (DM) interaction \cite{MoriyaT60, DzyaloshinskyI58} emerges, leading to a spiral magnetic order at its NESS. 

The rest of this paper is organized as follows. 
In Sec.~\ref{sec: FME, cla}, we present the general formalism on the FM expansion of a classical EOM with and without a stochastic variable. 
In Sec.~\ref{sec: conv}, we discuss the convergence property of the above FM expansion.
In Sec.~\ref{sec: Kapitza} and Sec.~\ref{sec: sLLG}, we illustrate our method in a single particle system and a many-body system, by examples of the Kapitza pendulum with friction and the periodically driven sLLG equation, respectively, and compare the time-periodic EOMs and their effective ones obtained by their FM expansion. 
In Sec.~\ref{sec: spintronics}, we present an application to spintronics, where we study a multiferroic spin chain irradiated by a laser. 
In Sec.~\ref{sec: summary}, we summarize the main results and discuss the outlook for future studies.

\section{\label{sec: FME, cla}FM expansion of a classical (stochastic) equation of motion}
\subsection{Equation of motion and master equation}
Consider a classical system under a periodic drive. 
Let $\bmphi(t) = \rbr{\phi_1(t), \phi_2(t), \cdots, \phi_N(t)}$ be a set of classical variables describing the system, e.g., the position of a particle for the Langevin equation and the magnetization for the sLLG equation, where $t$ denotes the time. 
We consider the system described by the following stochastic differential equation \cite{MorenoM15}: 
\begin{align}
	\dot{\phi}_i(t) = f_i\rbr{\bmphi(t), t}
	 + \sum_{j=1}^N g_{ij}\rbr{\bmphi(t), t} h_j(t)
	, \label{eq: sto, eom}
\end{align}
where $\dot{y} := dy/dt$ is the time derivative and $h_j$ is a Gaussian random variables that satisfies 
\begin{align}
	\langle h_i(t) \rangle &= 0
, \nonumber \\
	\langle h_i(t) h_j(t\pri) \rangle &= 2 D \delta_{ij}\delta(t - t\pri)
, \label{eq: sto, h}
\end{align}
with $D$ being the diffusion constant (the angle bracket $\langle \cdot \rangle$ denotes the average over different noise realizations). 
The drift force $f_i\br{\bmphi, t}$ and the diffusion matrix $g_{ij}\br{\bmphi, t}$ are time-periodic with period $T$: $f_i\br{\bmphi, t+T} = f_i\br{\bmphi, t}$ and $g_{ij}\br{\bmphi, t+T} = g_{ij}\br{\bmphi, t}$, and they are generally non-linear functions of $\bm{\phi}$. 
We note that when we consider an EOM for a classical field $\bmphi_{\bmr} = \rbr{\phi_{\bmr,1}(t), \phi_{\bmr,2}(t), \cdots, \phi_{\bmr,N_I}(t)}$, the subscript $i$ in Eq.~\eqref{eq: sto, eom} represents a collection of the coordinate $\bmr$ and the internal degrees of freedom $a$, where the EOM is written as follows: 
\begin{align}
	\dot{\phi}_{\bmr, a}(t) = 
	f_{\bmr, a}\rbr{\bmphi(t), t}
	 + \sum_{b = 1}^{N_I} g_{\bmr, ab}\rbr{\bmphi(t), t} h_{\bmr, b}(t)
	. \label{eq: sto, eom, field}
\end{align}
Here, we assume that the effect of the random field is local, i.e., the field $\bmphi_{\bmr}$ at $\bmr$ is affected by the random field $\bmh_{\bmr}$ at the same site $\bmr$. 
We choose the Stratonovich prescription for the application to the sLLG equation in Sec.~\ref{sec: sLLG} since the magnetization is conserved only in this prescription \cite{BertottiG09}. 
However, it is straightforward to generalize the following theory to the other prescriptions, e.g. the It$\mr{\hat{o}}$ and the post-point prescriptions \cite{MorenoM15}.

Equations \eqref{eq: sto, eom} and \eqref{eq: sto, eom, field} with a finite diffusion $D > 0$ are commonly used to describe diffusive processes in Nature, such as a Browninan motion \cite{KampenV92} and the magnetization dynamics of a micromagnet \cite{BrownW63}. 
For the overdamped Langevin equation in the one-dimensional space with potential $U(x)$ and mobility $\mu$, its EOM reads 
\begin{align}
	\dot{x} = - \mu {\partial U(x) \over \partial x} + h(t)
	,
\end{align}
where $\bmphi = x$ and $\bm{f} =- \mu \ \partial U(x) / (\partial x)$ are the position of the particle and the potential force, respectively, and $g_{ij} = 1$. 
When $D = 0$, Eq.~\eqref{eq: sto, eom} gives a deterministic equation
\begin{align}
	\dot{\phi}_i(t) = f_i\rbr{\bmphi(t), t}
	, \label{eq: fme, 10}
\end{align}
which describes an open classical system at sufficiently low temperature or an isolated classical one such as a Hamilton system. 

In addition to classical systems, our theory can be applied to even \textit{quantum} systems in symmetry-broken phases or semiclassical regimes because their EOMs take the form of Eqs.~\eqref{eq: sto, eom} or \eqref{eq: sto, eom, field}. 
In the former case, $\bmphi$ and Eq.~\eqref{eq: sto, eom} are the order parameter and its equation, e.g., the Gross-Pitaevskii (GP) equation \cite{PethickC02, KurnDS13} and the Ginzburg-Landau equation \cite{HohenbergP15}, respectively. 
For the GP equation, the order parameter $\bmphi = \tbr{\psi_{\bmr}}_{\bmr \in \mathbb{R}}$ represents the macroscopic wavefunction, with its equation written into the form of Eq.~\eqref{eq: sto, eom, field} with $D=0$:
\begin{align}
	\dot{\psi}_{\bmr} = 
	-i \rbr{ 
	- {\nabla_{\bmr}^2 \psi_{\bmr} \over 2m} 
	+ (\mu + g_c |\psi_{\bmr}|^2 )\psi_{\bmr}
	}
	. \label{eq: FME, GP}
\end{align}
Here, $m$, $\mu$ and $g_c$ are the mass of atoms, the chemical potential and the coupling constant, respectively. 
An example of the latter case is the Dicke model in the semiclassical limit \cite{BhaseenM12,BakemeierL13}, which describes two-level atoms coupled to a large number of photons in a cavity. 
In this limit, the system is described by the effective collective atomic pseudospin $\bm{J} := (J_x, J_y, J_z)$, which is a classical three-component vector, and the coherent-state amplitude $a$ of photons, which is a complex number. 
Its EOM is obtained from the Ehrenfest equation $d\langle A \rangle / dt = i d\langle \rbr{H,A} \rangle$ to be
\begin{align}
	\dot{\bm{J}} &= 
	(2 \lambda \mr{Re}\br{a} \hat{x} + \omega_a \hat{z})\times \bm{J}
	 , \nonumber \\
	\dot{a} &= -i \br{\omega_o a + \lambda J_x}
	 , 
\end{align}
where $\omega_a, \omega_o$, and $\lambda$ are the atomic frequency, the optical frequency, and the coupling constant between the photons and atoms, respectively, and $\hat{x} := (1,0,0)$ and $\hat{z} := (0,0,1)$.

By introducing the vector fields $\bm{f} := \br{f_1, f_2, \cdots, f_N}$ and $\bmh := \br{h_1, h_2, \cdots, h_N}$, and the matrix-valued function $G := \tbr{g_{ij}}_{i,j=1}^N$, 
we can rewrite Eqs.~\eqref{eq: sto, eom} and \eqref{eq: sto, h} in compact forms:  
\begin{align}
	& \dot{\bmphi} = \bm{f}(\bmphi, t) + G(\bmphi, t)\bmh(t)
	, \nonumber \\
	& \left< \bmh(t) \right> = 0
	, \nonumber \\
	& \left< \rbr{\bmh(t)}^{\mr{tr}} \bmh(t\pri) \right> = 2 D I_N \delta(t - t\pri)
	, \label{eq:matrixEOM}
\end{align}
where the superscript $\mr{tr}$ denotes the transpose. 
Using the standard technique of the stochastic calculus \cite{BrownW63, CoffeyT04, PalaciosJ98, BertottiG09}, 
we obtain the master equation, which is known as the Fokker-Planck equation, corresponding to Eq.~\eqref{eq: sto, eom}: 
\begin{align}
	{\partial P(\bmphi,t) \over \partial t} 
	&= 
	{\partial \over \partial \phi_i} \left[
	\mathcal{F}_i(\bmphi,t) P(\bmphi,t)
	\right]
	\ \nonumber \\
	& \quad + {\partial^2 \over \partial \phi_i \partial \phi_j} \left[
	\mathcal{D}_{ij}(\bmphi,t) P(\bmphi,t)
	\right]
	, \label{eq: fokker}
\end{align}
where 
\begin{align}
	\mathcal{F}_i(\bmphi,t) &:= - f_i(\bmphi,t) - D g_{kl}(\bmphi,t) {\partial g_{il}(\bmphi,t) \over \partial \phi_k}
	, \nonumber \\
	\mathcal{D}_{ij}(\bmphi,t) &:= D g_{ik}(\bmphi,t) g_{jk}(\bmphi,t)
	. \label{eq: fokker, coef} 
\end{align}
Here, $P(\bmphi\pri,t)$ is the probability density for finding the variable $\bmphi = \bmphi\pri$ at time $t$ in the whole parameter space of $\bmphi$, and we omit the summation of the repeated indices $i, j, k, l$. 
Equations \eqref{eq:matrixEOM}, \eqref{eq: fokker}, and \eqref{eq: fokker, coef} give the correspondence between a EOM and a master equation. 
By introducing the current $J_i := -\mathcal{F}_i P - \partial (D_{ij} P)/ (\partial \phi_j)$, we can rewrite Eq.~\eqref{eq: fokker} into the continuity equation for $P$:
\begin{align}
	\partial_t P + \mr{div}{\bm{J}} = 0
	, 
\end{align}
where $\mr{div}{\bm{J}}:=\sum_i\partial J_i/\partial \phi_i$. 
This equation satisfies the conservation of the probability: $\int d\bm{\phi} P(\bm{\phi}, t) = 1$ for $\forall t$. 
We note that the master equation \eqref{eq: fokker} contains only up to the second-order derivative of $\bm{\phi}$ because $\bm{h}$ is a Gaussian random variable with a Markovian nature. 
In other words, if the master equation contains higher-order derivative or becomes an integro-differential equation, the random noise must be either non-Markovian or non-Gaussian \cite{KanazawaK12, KanazawaK13, ZwanzigR61, MoriH73, HaakeF73}. 
We will again comment on this issue in the next subsection. 

If we introduce the vector field $\bm{\mathcal{F}} := (\mathcal{F}_1, \mathcal{F}_2, \cdots, \mathcal{F}_N)$ and the matrix-valued field $\mathcal{D} := \tbr{\mathcal{D}_{ij}}_{i,j=1}^N$, 
Eq.~\eqref{eq: fokker} can be rewritten in a compact form:
\begin{align}
	\partial_t P(\bmphi,t)
	 &= 
	\mr{div}\br{ \bm{\mathcal{F}} P }
	 + \mr{div}_2\br{ \mathcal{D} P }
	,
\label{eq:materixMaster}
\end{align}
where the operator $\mr{div}_2$ on a matrix $\mathcal{D}\pri = \tbr{\mathcal{D}_{ij}(\bmphi)}_{ij}$ is defined by $\mr{div}_2(\mathcal{D}\pri) := (\partial^2 \mathcal{D}\pri_{ij}) / (\partial \phi_i \partial \phi_j)$.

\subsection{FM expansion of a master equation}
\begin{figure}[t!]
\centering
   \includegraphics[width=\columnwidth, clip]{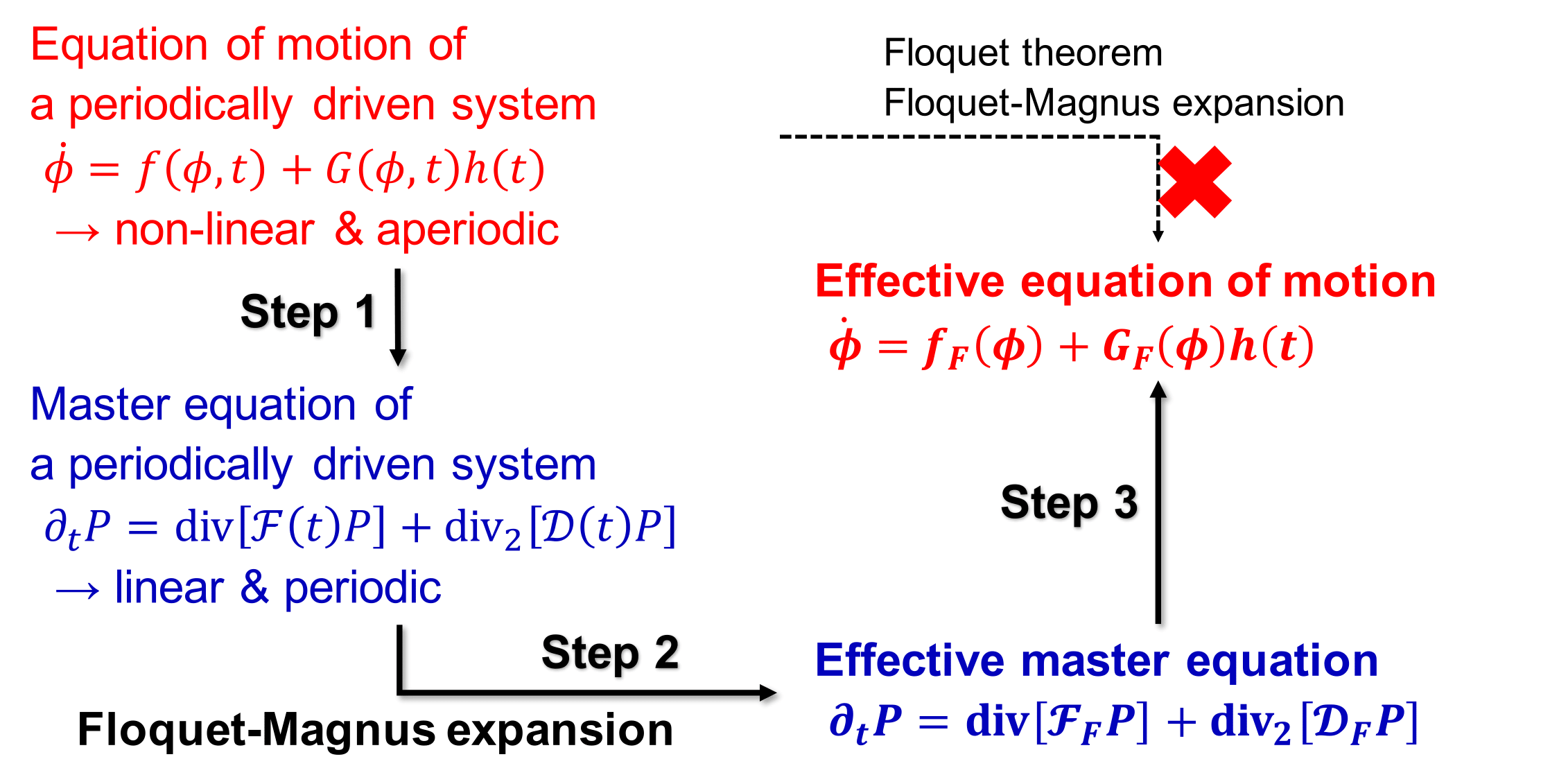}
\caption{Procedure for obtaining the FM expansion of the classical EOM described by a stochastic differential equation \eqref{eq: sto, eom}. 
	In the first step, we turn to the master equation \eqref{eq: fokker} corresponding to Eq.~\eqref{eq: sto, eom}, to which we perform the FM expansion to obtain the effective master equation \eqref{eq: FME, eff, master} in the second step.  
	Finally, we find a stochastic differential equation corresponding to Eq.~\eqref{eq: FME, eff, master} in the third step. 
    }
       \label{fig: strategy}
\end{figure}  

In an analysis of a driven quantum system, Floquet theorem \cite{FloquetG83, ChiconeC06} and the FM expansion \cite{GoldmanN14a, EckardtA15, BukovM15} are commonly used techniques because they allow us to map a non-equilibrium system to an effective static one, thereby simplifying its analysis. 
Unfortunately, we cannot apply them directly to classical EOMs because the original equation \eqref{eq: sto, eom} is neither linear in $\bmphi$ nor time-periodic due to the random variable $h_j(t)$. 
Nevertheless, we can apply it to its master equation \eqref{eq: fokker} because it is linear in $P$ and time-periodic. 
Figure~\ref{fig: strategy} summarizes our strategy. 
The FM expansion of an EOM is performed via that of the corresponding master equation. 

By introducing the Fokker-Planck operator $\mL_t$ defined by 
\begin{align}
	\mL_t(P) &= 
	\mr{div}\rbr{ \bm{\mathcal{F}}(t) P }
	 + \mr{div}_2\rbr{ \mathcal{D}(t) P }
	, 
\end{align}
we can regard Eq.~\eqref{eq: fokker} as the ``Schr\"odinger equation" driven by the non-Hermitian time-periodic ``Hamiltonian" $H(t) := i \mL_t$ \cite{HinrichsenH00}: 
\begin{align}
	i \partial_t P = H(t)P
	, 
	\label{eq: periodic, sch}
\end{align}
where the probability distribution $P$ plays the role of a wave function. 
We can formally solve Eq.~\eqref{eq: periodic, sch} as $P(\bmphi, t) = U(t, 0) P(\bmphi, t=0)$, where $U(t_2, t_1)$ is the time evolution operator from $t_1$ to $t_2$. 
From the Floquet theorem, we can rewrite $U(t_2, t_1)$ using the effective Hamiltonian $H_F$ and an anti-hermitian operator $\mG_F(s)$ as follows \cite{EckardtA15, BukovM15}: 
\begin{align}
	U(t_2, t_1) &:= \mathcal{T}_{t} \exp \rbr{
	- i \int_{t_1}^{t_2} H(t) dt
	}
	\nonumber \\
	 & = \Exp{\mG_F(t_2)}
	 \Exp{ (t_2 - t_1) \mL_F} 
	 \Exp{- \mG_F(t_1)}
	 , \label{eq: time, evo, Floquet}
\end{align}
where $\mathcal{T}_{t}$ is the time ordering operator and $\mL_F := -i H_F$ is the effective Fokker-Planck operator.  
In the context of quantum systems, $i \mG_F(s)$ is known as the kick operator \cite{GoldmanN14a} and satisfies the time-periodicity $\mG_F(s + T) = \mG_F(s)$, which represents an instantaneous time evolution at time $s$ and induces a small oscillational motion around the slower dynamics. 
In what follows, we call $\mG_F(t)$ itself the kick operator as well. 
Note that the Floquet theorem itself does not require the hermiticity of $H(t)$ and hence apply to the non-Hermitian Schr\"odinger equation \eqref{eq: periodic, sch}. 

For a fast drive, the effective Fokker-Planck operator $\mL_F$ and the kick operator $\mG_F(t)$ can be formally expanded in powers of $\omega^{-1}$ as follows \cite{PratoD97, BialynickiI69, GoldmanN14a}:
\begin{align}
	\mL_F &:= \sum_{m=0}^\infty \mL_F^{(m)}
	, \nonumber \\ 
	\mG_F(t) &:= \sum_{m=0}^\infty \mG_F^{(m)}(t)
	, 
\end{align}
where $\mL_F^{(m)}$ and $\mG_F^{(m)}$ are of the order of $\omega^{-m}$. 
As we will show below, $\mL_F^{(m)}$ and $\mG_F^{(m)}$ are derived in the same manner as done for quantum systems \cite{GoldmanN14a, EckardtA15}. 

Let us expand $\mL_t$ and $H(t)$ in their Fourier harmonics as follows: 
\begin{align}
	\mL_t & = \sum_m \mL_m \Exp{-im\omega t}
	, \nonumber \\
	H(t) & = \sum_m H_m \Exp{-im\omega t}
	, 
\end{align}
where $\mL_m = - i H_m$. 
Conversely, $\mL_m$ ($H_m$) is determined from $\mL_t$ ($H(t)$) as $\mL_m  = {1 \over T} \int_0^T dt \mL_t \Exp{im\omega t}$ ($H_m = {1 \over T} \int_0^T dt H(t) \Exp{im\omega t}$). 
The first three terms of $\mL_F^{(m)}$ are expressed as follows \cite{GoldmanN14a, EckardtA15}: 
\begin{align}
	\mL_F^{(0)}
	& = -i H_F^{(0)}
	 = -i H_0 = \mL_0
	, \nonumber \\
	\mL_F^{(1)}
	& = -i H_F^{(1)} 
	 = -i \sum_{m \ne 0}
	{\rbr{H_{-m}, H_m} \over 2 m \omega}
	, \nonumber \\
	& = i \sum_{m\ne 0}
	{\rbr{\mL_{-m}, \mL_m} \over 2m \omega}
	, \nonumber\\
	\mL_F^{(2)} 
	& = -i H_F^{(2)}
	\nonumber \\
	& = -i \sum_{m\ne 0} 
	\tbr{{\rbr{H_{-m}, \rbr{H_0, H_m}} 
	\over 2(m \omega)^2} \right.
	\nonumber \\
	&\left. \quad \quad + \sum_{m\pri \ne 0, m} 
	 {\rbr{H_{-m\pri}, \rbr{H_{m\pri - m}, H_m}} 
	 \over 3m m\pri \omega^2}}
	, \nonumber \\
	& = i^2 \sum_{m\ne 0} 
	\tbr{{\rbr{\mL_{-m}, \rbr{\mL_0, \mL_m}} 
	\over 2(m \omega)^2} \right.
	\nonumber \\
	&\left. \quad \quad + \sum_{m\pri \ne 0, m} 
	 {\rbr{\mL_{-m\pri}, \rbr{\mL_{m\pri - m}, \mL_m}} 
	 \over 3m m\pri \omega^2}}
	, \label{eq: fm, fokker, 11}
\end{align}
where the kick operators $\mG_F^{(m)}$ are obtained to be  
\begin{align}
	\mG_F^{(0)}(t)
	& = 0
	, \nonumber \\
	\mG_F^{(1)}(t)
	& = - \sum_{m \ne 0}
	{H_{-m} \Exp{im \omega t} \over m \omega}
	 = - i \sum_{m \ne 0}
	{\mL_{-m} \Exp{im \omega t} \over m \omega}
	, \nonumber \\
	\mG_F^{(2)}(t)
	& = \sum_{m\ne 0} 
	\left\{{\rbr{H_0, H_{-m}} \Exp{i m \omega t}
	\over (m \omega)^2} \right.
	\nonumber \\
	&\left. \quad + 
	\sum_{m\pri \ne 0, m} 
	 {\rbr{H_{m\pri}, H_{-m}} \Exp{i (m - m\pri) \omega t}
	 \over 2 m (m - m\pri) \omega^2} \right\}
	\nonumber \\
	& = i^2 \sum_{m\ne 0} 
	\left\{{\rbr{\mL_0, \mL_{-m}} \Exp{i m \omega t}
	\over (m \omega)^2} \right.
	\nonumber \\
	&\left. \quad + 
	\sum_{m\pri \ne 0, m} 
	 {\rbr{\mL_{m\pri}, \mL_{-m}} \Exp{i (m - m\pri) \omega t}
	 \over 2 m (m - m\pri) \omega^2} \right\}
	. \label{eq: fm, fokker, 12}
\end{align} 
We note that the commutator $\rbr{\cdot, \cdot}$ in Eqs.~\eqref{eq: fm, fokker, 11} and \eqref{eq: fm, fokker, 12} is interpreted as that between operators: 
\begin{align}
	\rbr{\mathcal{S}_1, \mathcal{S}_2}(P) := 
	\mathcal{S}_1\rbr{\mathcal{S}_2(P)} - 
	\mathcal{S}_2\rbr{\mathcal{S}_1(P)}
	. \label{eq: comm, ope}
\end{align}
We further note that we have taken the convention $\int_0^T \mathcal{G}_F(t) dt = 0$ such that the effective Fokker-Planck operator becomes time-independent. 
For quantum systems in this convention, $\mathcal{G}_F$ and $\mL_F$ are the same as those obtained from the van Vleck degenerate perturbation theory \cite{EckardtA15, BukovM15, ManangaE11}. 

In summary, the effective Fokker-Planck operator $\mL_F^{(m)}$ is obtained by formally replacing the $m$th coefficient $H_{m}$ of the Hamiltonian by $\mL_{m}$, the $m$th coefficient of the Fokker-Planck operator, and the commutator between the Hamiltonians by the commutator \eqref{eq: comm, ope}, and finally multiplying $i^m$. 
If we focus on the effective dynamics ignoring the kick operator, the master equation is given by the following time-independent equation   
\begin{align}
	\partial_t P = \mL_F P \approx \sum_{m=0}^{m_0} \mL_F^{(m)} P 
	, \label{eq: FME, eff, master}
\end{align}
where $m_0$ is the truncation order.
So far, our argument has focused only on the FM expansion. However, it is straightforward to generalize above analysis to other expansions like the Brillouin-Wigner expansion \cite{MikamiT16} and the Floquet-Schriefer-Wolff transformation \cite{MentinkJ15, BukovM16, KitamuraS17, ClaassenM17}. 

To complete the procedure in Fig.~\ref{fig: strategy}, we must find an EOM whose master equation coincides with the truncated effective master equation obtained from the FM expansion (the step 3 in Fig.~\ref{fig: strategy}). 
In general, this problem is nontrivial. 
When the master equation includes at most the second-order derivative terms $\partial^2(\mathcal{D}_{ij}P)/ (\partial \phi_i \partial \phi_j)$, we can find the corresponding stochastic differential equation through Eq.~\eqref{eq: fokker, coef}. 
As we will see below, several physically relevant situations are included in this case such as the systems without a diffusion ($\mathcal{D} = 0$) or a time-independent diffusion ($\mathcal{D}(t) = \mathrm{const}$).
On the other hand, when the truncated FM expansion of a master equation contains derivatives higher than the second-order (e.g., $\partial^3(\mathcal{D}_{ijk}P)/ (\partial \phi_i \partial \phi_j \phi_k)$), the random variable $h_j$ must be either non-Markovian or non-Gaussian \cite{KanazawaK12, KanazawaK13, ZwanzigR61, MoriH73, HaakeF73}. 
In fact, for the case a quantum master equation under a periodic drive, non-Markov properties appear in its effective dynamics, which is well captured by introducing a memory kernel \cite{SchnellA18}. 
It is unclear whether one can find an appropriate random variable with non-Markovian or non-Gaussian properties even in our problems, though approximation schemes have been developed \cite{KanazawaK15, RiskenH96}.

\subsection{\label{sec: FME, cla, det}Deterministic system}
We here consider the system without a diffusion, i.e., $\mathcal{D} = 0$ and $h_j(t) = 0$. 
In this case, its EOM in Eq.~\eqref{eq: fme, 10} can be regarded as a flow equation generated by $\bm{f}(\bmphi, t)$. 
Let us define $\bm{f}_m$ as the $m$th-order Fourier harmonics of $\bm{f}(\bmphi, t)$: 
\begin{align}
	\bm{f}(\bmphi, t) = \sum_m \bm{f}_m \Exp{-im\omega t}
	. 
\end{align}
Then, $\mL_m$ is written as $\mL_m(P) := - \mr{div}(\bm{f}_m P)$, and the commutator $\rbr{\mL_m, \mL_n}$ is obtained to be 
\begin{align}
	\rbr{\mL_m, \mL_n}(P) &= 
	\mr{div}\rbr{\bm{f}_m \mr{div}(\bm{f}_n P)} - 
	\mr{div}\rbr{\bm{f}_n \mr{div}(\bm{f}_m P)}
	\nonumber \\
	 & =
	\mr{div}\tbr{\rbr{
	\br{\bm{f}_m \cdot \nabla_{\bm\phi}}\bm{f}_n - 
	\br{\bm{f}_n \cdot \nabla_{\bm\phi}}\bm{f}_m
	}P}
	\nonumber \\
	 &= 
	- \mr{div}\br{
	 - \rbr{\bm{f}_m, \bm{f}_n}_{\mr{cl}}
	P}
	. \label{eq: comm, closed}
\end{align}
Here $\nabla_{\bm\phi}:=(\partial/\partial\phi_1, \partial/\partial\phi_2,\cdots)$ and the commutator $\rbr{\bm{A}, \bm{B}}_{\mr{cl}}$ between two vector fields $\bm{A}$ and $\bm{B}$ is defined by 
\begin{align}
	\rbr{\bm{A}, \bm{B}}_{\mr{cl},j} & = 
	\br{\bm{A} \cdot \nabla_{\bm\phi}}B_j - 
	\br{\bm{B} \cdot \nabla_{\bm\phi}}A_j
	\nonumber \\
	& := A_i {\partial B_j \over \partial \phi_i}
	 - B_i {\partial A_j \over \partial \phi_i}
	, 
\end{align}
which is called the Lie bracket in mathematics.  
Equation \eqref{eq: comm, closed} implies that the operators of the form $\mL := \mr{div}(\bm{f} \cdot)$ is closed with respect to the commutator, and thereby the effective dynamics is described by the renormalized drift force $\bm{f}_F$. 
The $m$th-order FM expansion $\bm{f}_F^{(m)}$ of the drift field is obtained from that of the effective Hamiltonian by replacing the commutator $\rbr{H_m, H_n}$ between Hamiltonians with $\rbr{\bm{f}_m, \bm{f}_n}_{\mr{cl}}$, i.e., the commutator between drift field, followed by the multiplication by $i^m$. 
Then, the resulting effective EOM up to the order of $\omega^{-2}$ is obtained to be 
\begin{align}
	\dot{\bmphi} &= \bm{f}_F(\bmphi)
	\nonumber \\ 
	& = \bm{f}_0(\bmphi)
	 + i \sum_{m \ne 0}
	{ \rbr{\bm{f}_{-m}, \bm{f}_m}_{\mr{cl}} \over 2 m \omega}
	\nonumber \\
	 & - \sum_{m\ne 0} 
	\tbr{{\rbr{\bm{f}_{-m}, \rbr{\bm{f}_0, \bm{f}_m}_{\mr{cl}}}_{\mr{cl}}
	\over 2(m \omega)^2} \right.
	\nonumber \\
	&\left. \quad \quad + \sum_{m\pri \ne 0, m} 
	 {\rbr{\bm{f}_{-m\pri}, 
	 \rbr{\bm{f}_{m\pri - m}, \bm{f}_m}_{\mr{cl}}}_{\mr{cl}} 
	 \over 3m m\pri \omega^2}}
	  + \mathcal{O}(\omega^{-3})
	. \label{eq: FME, deterministic}
\end{align} 
This result is consistent with the Magnus expansion of general non-autonomous (not necessarily time-periodic) ordinary differential equation $\dot{\bmphi} = \bm{f}(\bmphi, t)$ \cite{BlanesS09, SpirigF79, AgrachevA81, GamkrelidzeA79}. 
As a special case, if the dynamics are governed by some classical Hamiltonian $H(t)$, the drift field $\bm{f}(t)$ and the commutator $\rbr{\cdot, \cdot}_{\mr{cl}}$ are replaced by the Hamilton flow and the Poisson bracket $- \tbr{\cdot, \cdot}$, respectively. 
Note that the master equation \eqref{eq: fokker} is nothing but the Liouville equation. 
Let $\bm{q}$ and $\bm{p}$ be a canonical conjugate pair. 
Then, the classical variable $\bmphi$ and $\bm{f}_m$ are given by 
\begin{align}
	\bmphi &= (\bm{q}, \bm{p})
	, \quad
	\bm{f}_m = 
	\br{
	{\partial H_m \over \partial \bm{p}}, 
	 - {\partial H_m \over \partial \bm{q}}
	}
	. 
\end{align}
By a straightforward calculation, we obtain 
\begin{align}
	\rbr{\bm{f}_m, \bm{f}_n}_{\mr{cl}} = 
	- \br{
	{\partial \tbr{H_m, H_n} \over \partial \bm{p}}, 
	 - {\partial \tbr{H_m, H_n} \over \partial \bm{q}}
	}
	. 
\end{align}
The above results correctly reproduce the previous ones for periodically driven isolated Hamilton systems \cite{MoriT18, BukovM15, DAlessioL13} and are consistent with the Magnus expansion of general time-dependent (not necessarily time-periodic) Hamilton systems \cite{RosJ91, TaniS68, MarcusR70}.

\subsection{Time-independent diffusion}
We here assume that the diffusion matrix $G$ and hence $\mathcal{D}$ are time-independent, where the $\mL_m$ is given by  
\begin{align}
	\mL_0(P) &:= 
	\mr{div}\br{\bm{\mathcal{F}}_{0} P} + 
	\mr{div}_2 \rbr{\mathcal{D} P }
	, \nonumber \\
	\mL_m(P) &:= 
	\mr{div}\br{\bm{\mathcal{F}}_m P}
	\quad \mr{for} \quad m \ne 0
	.
\end{align}
Here, $\bm{\mathcal{F}}_m$ is the $m$th-order Fourier harmonics of $\bm{\mathcal{F}}(t)$: $\bm{\mathcal{F}}(t) = \sum_m \bm{\mathcal{F}}_m \Exp{-im\omega t}$. 
Under this assumption, one can always find the effective EOM corresponding to the effective Fokker-Planck operator $\mL_F$ if we truncate at the second order.  
From Eq.~\eqref{eq: fm, fokker, 11}, the first order is obtained to be 
\begin{align}
	\mL_F^{(1)} &:= 
	 i \sum_{m \ne 0}  
	{\rbr{\mL_{-m}, \mL_m} \over 2m \omega}
	 = \mr{div} \rbr{
	\bm{\mathcal{F}}_F^{(1)} \cdot
	}
	, \nonumber \\
	\bm{\mathcal{F}}_F^{(1)} &:= 
	\sum_{m \ne 0} 
	{i \over 2m \omega} \rbr{
	\br{\bm{\mathcal{F}}_{-m} \cdot \nabla_{\bm\phi}}\bm{\mathcal{F}}_m
	 - \br{\bm{\mathcal{F}}_m \cdot \nabla_{\bm\phi}}\bm{\mathcal{F}}_{-m}
	} 
	\nonumber \\
	 &= \sum_{m \ne 0} 
	 {i \over 2m \omega} \rbr{
	\br{\bm{f}_{-m} \cdot \nabla_{\bm\phi}}\bm{f}_m
	 - \br{\bm{f}_m \cdot \nabla_{\bm\phi}}\bm{f}_{-m}
	} 
	, 
\end{align}
where the effective Fokker-Planck operator is given by 
\begin{align}
	\mL_F(P) &= 
	\mr{div}\rbr{(\bm{\mathcal{F}}_{0} + \bm{\mathcal{F}}_F^{(1)}) P}
	 + \mr{div}_2 (\mathcal{D} P)
	. \label{eq: Fokker-Planck, 1st order}
\end{align}
A crucial observation here is that only the drift field is renormalized in the first order. 
Therefore, we can always find the EOM with Fokker-Planck operator \eqref{eq: Fokker-Planck, 1st order}, which is obtained to be
\begin{align}
	\dot{\bmphi} = \bm{f}_F(\bmphi) + G(\bmphi) h(t)
	, \label{eq: FME, 1st, evo}
\end{align}
where $\bm{f}_F$ is the renormalized drift term:  
\begin{align}
	\bm{f}_F := \bm{f}_0 + 	
	\sum_{m \ne 0} {i \over 2m \omega} \rbr{
	(\bm{f}_m \cdot \nabla_{\bm\phi}) \bm{f}_{-m}
	 - (\bm{f}_{-m} \cdot \nabla_{\bm\phi}) \bm{f}_m
	}
	. \label{eq: FME, 1st}
\end{align}
Notably, when $\bmf_F$ represents a potential force with a renormalized potential $V_F(\bmphi)$ and $G(\bmphi)$ satisfies the detailed-balance condition with temperature $T_{\mr{te}}$, the system thermalizes into a canonical distribution $\exp(-V_F(\bmphi) / T_{\mr{te}})$ of the potential $V_F(\bmphi)$. 
The time evolution without the kick operator is obtained by simply solving Eq.~\eqref{eq: FME, 1st, evo}.

On the other hand, the calculation of the time evolution  with the kick operator is constituted from three steps corresponding to the three exponential operators $\exp\rbr{- \mG_F(t_1)}$, $\exp\rbr{ (t_2 - t_1) \mL_F}$, and $\exp\rbr{\mG_F(t_2)}$, as shown in Eq.~\eqref{eq: time, evo, Floquet}. 
Let us first consider the effect of the kick operators on the EOM.
The kick operator $\mG_F^{(1)}(s)$, with $s \ (= t_1, t_2)$ being either the initial or final kick time, 
is given from Eq.~\eqref{eq: fm, fokker, 12} as  
\begin{align}
	\mG_F^{(1)}(s) &= - {1 \over \omega} 
	\mr{div}\rbr{\bmf_{F, \mr{mic}}^{(1)}(\bmphi, s) \ \ \cdot \ \ }, 
\end{align}
where $\bmf_{F, \mr{mic}}^{(1)}$ is the oscillating drift field: 
\begin{align}	
	\bmf_{F, \mr{mic}}^{(1)}(\bmphi, s) &= 
	-i \sum_{m \ne 0} 
	{\bm{f}_{-m} \Exp{i \omega s} \over 2m} 
	. \label{eq: kick, 1st}
\end{align}
From Eq.~\eqref{eq: kick, 1st} and the definition of an exponential operator, $\exp\rbr{\pm \mathcal{G}_F(s)}P_0(\bmphi)$ is formally the solution of the master equation 
\begin{align}
	{\partial P(\bmphi, \tau) \over \partial \tau} &= 
	\mp \mr{div}\rbr{\bmf_{F,\mr{mic}}^{(1)}(\bmphi, s) P(\bmphi, \tau)}
	, \nonumber \\
	P(\bmphi, \tau=0) &= P_0(\bmphi)
	, \label{eq: kick, equation1}
\end{align}
at time $\tau = 1/\omega$. 
Thus, the exponential operator $\exp\rbr{\pm \mathcal{G}_F(s)}$ has the following physical interpretation: 
it is the integration of the flow field $\bmf_{F, \mr{mic}}^{(1)}$ from $\tau = 0$ to $\tau = 1/\omega$, where $\tau$ is an auxiliary time for calculating the kicks and $1/\omega$ is the duration of the kick. 
Since $\mG_F^{(1)}$ does not contain the diffusion term, plugging $P_0(\bmphi_{\mr{kick}}) = \delta(\bmphi_{\mr{kick}} - \bmphi_0)$ into Eq.~\eqref{eq: kick, equation1}, we can rewrite it into the equation for $\bmphi_{\mr{kick}}$: 
\begin{align}
	{d \bmphi_{\mr{kick}} \over d \tau}
	&= \pm \bmf_{F,\mr{mic}}^{(1)}(\bmphi_{\mr{kick}}, s) , \quad
	\bmphi_{\mr{kick}}(\tau=0) = \bmphi_0
	. \label{eq: kick, equation2}
\end{align}
Thus, $\bmphi_0$ is mapped to the solution of Eq.~\eqref{eq: kick, equation2} at time $\tau = 1/\omega$ by the initial kick. 
Note that Eqs.~\eqref{eq: kick, equation1} and \eqref{eq: kick, equation2} are autonomous equations, i.e., they do not explicitly depend on the time $\tau$. 
Practically, due to the smallness of the integration time $1/\omega$, the solution $\bmphi_{\mr{kick}}(\tau = 1/\omega)$ is well approximated by the Euler method:
\begin{align}
	\bmphi_{\mr{kick}}
	\br{\tau = { 1 \over \omega}} 
	\simeq
	\bmphi_0 \pm {\bmf_{F, \mr{mic}}^{(1)} (\bmphi_0, s) \over \omega}
	. 
\end{align} 

We now see the generic effect of the kick operator, and then the three-step computation of the time evolution operator 
$U(t_2,t_1)=\Exp{\mG_F(t_2)}\Exp{ (t_2 - t_1) \mL_F}\Exp{-\mG_F(t_1)}$ is performed as follows. 
Let $\bmphi_0$ be the initial state of $\bmphi$.
First, to calculate the effect of the initial kick $\exp\rbr{- \mG_F(t_1)}$, we integrate Eq.~\eqref{eq: kick, equation2} with the minus sign on the right-hand side and $s = t_1$ for the initial state $\bmphi_0$:
\begin{align}
	{d \bmphi_{\mr{kick}} \over d \tau}
	&= - \bmf_{F,\mr{mic}}^{(1)}(\bmphi_{\mr{kick}}, t_1) , \quad
	\bmphi_{\mr{kick}}(\tau=0) = \bmphi_0
	. 
\end{align}
The solution at time $\tau = 1/\omega$ gives the state after the initial kick, which we write as $\bmphi_1$: $\bmphi_1 := \bmphi_{\mr{kick}}(\tau = 1/\omega)$. 
Next, we evaluate the effective dynamics $\exp\rbr{ (t_2 - t_1) \mL_F}$ by integrating Eq.~\eqref{eq: FME, 1st, evo} from within time $(t_2 - t_1)$ with the initial state taken as $\bmphi_1$: 
\begin{align}
	{d\bmphi_{\mr{eff}} \over dt} = \bm{f}_F(\bmphi_{\mr{eff}}) + G(\bmphi_{\mr{eff}}) h(t)
	, \quad
	\bmphi_{\mr{eff}}(t = 0) = \bmphi_1
	. 
\end{align}
The solution at $t = t_2 - t_1$ gives the state after the effective flow $\mL_F$, which we write as $\bmphi_2$: $\bmphi_2 := \bmphi_{\mr{eff}}(t = t_2 - t_1)$. 
Finally, we integrate Eq.~\eqref{eq: kick, equation2} with the plus sign on the right-hand side and $s = t_2$ up to time $1/\omega$ for the initial state $\bmphi_2$ to calculate the final kick: 
\begin{align}
	{d \bmphi_{\mr{kick}} \over d \tau}
	&= \bmf_{F,\mr{mic}}^{(1)}(\bmphi_{\mr{kick}}, t_2) , \quad
	\bmphi_{\mr{kick}}(\tau = 0) = \bmphi_2
	. 
\end{align}
Then, the solution $\bmphi_3 = \bmphi_{\mr{kick}}(\tau = 1/\omega)$ gives the state after the final kick, 
and hence the state applied by the three operators $\Exp{-\mG_F(t_1)}$, $\Exp{ (t_2 - t_1) \mL_F}$, and $\Exp{\mG_F(t_2)}$ to the initial state $\bmphi_0$.

In the calculation at the second order, there appears the commutator between $\mL(P) = \mr{div}(\bm{\mathcal{F}} P) + \mr{div}_2(\mathcal{D} P)$ and $\mL\pri(P) = \mr{div}(\bm{\mathcal{F}}\pri P)$, which is calculated as follows: 
\begin{align}
	& \rbr{\mL, \mL\pri}(P)
	\nonumber \\
	& = 
	\mr{div}\rbr{\bm{\mathcal{F}} \mr{div}(\bm{\mathcal{F}}\pri P)}
	 - \mr{div}\rbr{ \bm{\mathcal{F}}\pri \mr{div}(\bm{\mathcal{F}} P)}
	 \nonumber \\
	&\quad + \mr{div}_2\rbr{\mathcal{D} \mr{div}(\bm{\mathcal{F}}\pri P)}
	 - \mr{div}\rbr{\bm{\mathcal{F}}\pri \mr{div}_2(\mathcal{D} P)}
	 \nonumber \\
	 & =: 
	\mr{div}\rbr{\mr{drf}(\bm{\mathcal{F}}, \bm{\mathcal{F}}\pri, \mathcal{D}) P}
	  + \mr{div}_2\rbr{\mr{diff}(\bm{\mathcal{F}}\pri, \mathcal{D}) P}
	  , 
\end{align}
where the corresponding drift field $\mr{drf}(\bm{\mathcal{F}}, \bm{\mathcal{F}}\pri, \mathcal{D}) := \tbr{\mr{drf}_i(\bm{\mathcal{F}}, \bm{\mathcal{F}}\pri, \mathcal{D})}_{i=1}^N$ and the diffusion matrix $\mr{diff}(\bm{\mathcal{F}}\pri, \mathcal{D}) = \tbr{ \mr{diff}_{ij}(\bm{\mathcal{F}}\pri, \mathcal{D}) }_{i,j =1}^N$ are given by 
\begin{align}
	\mr{drf}_i(\bm{\mathcal{F}}, \bm{\mathcal{F}}\pri, \mathcal{D}) &= 
	(\bm{\mathcal{F}} \cdot \nabla_{\bm\phi})\mathcal{F}\pri_i 
  - (\bm{\mathcal{F}}\pri \cdot \nabla_{\bm\phi})\mathcal{F}_i
	 - {\partial^2 \mathcal{F}\pri_i \over \partial \phi_j \partial \phi_k} \mathcal{D}_{jk}
	, \nonumber \\
	\mr{diff}_{ij}(\bm{\mathcal{F}}\pri, \mathcal{D}) & =
	{ \partial \mathcal{F}\pri_i \over \partial \phi_k} \mathcal{D}_{kj}
	 + { \partial \mathcal{F}\pri_j \over \partial \phi_k} \mathcal{D}_{ki}
	 - \mathcal{F}\pri_k { \partial \mathcal{D}_{ij} \over \partial \phi_k} 
	. \label{eq: FME, 2nd, 1}
\end{align}
The second-order term $\mL_F^{(2)}$ is obtained to be
\begin{align}
	\mL_F^{(2)}(P) &= \mr{div}(\bm{\mathcal{F}}_F^{(2)} P) + \mr{div}_2(\mathcal{D}_F^{(2)} P)
	, \nonumber \\
	\bm{\mathcal{F}}_F^{(2)} &:= 
	\sum_{m\ne 0} 
	\left\{{
	\mr{drf}\rbr{\mr{drf}\rbr{
	\bm{\mathcal{F}}_0, \bm{\mathcal{F}}_m, \mathcal{D}}, \bm{\mathcal{F}}_{-m}, \mathcal{D}}
	 \over 2(m \omega)^2
	 }\right. \nonumber \\
	 & \quad \quad + \sum_{m\pri \ne 0, m} 
	 \left.
	 {-\rbr{ \bm{\mathcal{F}}_{-m\pri}, \rbr{\bm{\mathcal{F}}_{m-m\pri}, \bm{\mathcal{F}}_m}_{\mr{cl}}}_{\mr{cl}}
	\over 3m m\pri \omega^2} \right\}
	, \nonumber \\
	\mathcal{D}_F^{(2)} &:= 	
	\sum_{m\ne 0} 
	{\mr{diff}\rbr{\bm{\mathcal{F}}_{-m}, 
	\mr{diff}\br{\bm{\mathcal{F}}_m, \mathcal{D}}
	} 
	\over 2(m \omega)^2}
	, \label{eq: FME, 2nd, 2}
\end{align}
which indicates that not only the drift vector but also the diffusion matrix is renormalized at the second order. 
The kick operator $\mG_F^{(2)}$ is given by 
\begin{align}
	\mG_F^{(2)}(t) &= \mr{div}(\bm{\mathcal{F}}_{F, \mr{mic}}^{(2)} \cdot) + \mr{div}_2(\mathcal{D}_{F, \mr{mic}}^{(2)} \cdot)
	, \nonumber \\
	\bm{\mathcal{F}}_{F, \mr{mic}}^{(2)} &= 
	- \sum_{m\ne 0} 
	\left\{{
	\mr{drf}\rbr{\bm{\mathcal{F}}_0, \bm{\mathcal{F}}_{-m}, \mathcal{D}} \Exp{i m \omega t}
	\over (m \omega)^2} \right.
	\nonumber \\
	&\left. \quad + 
	\sum_{m\pri \ne 0, m} 
	 {\rbr{\bm{\mathcal{F}}_{m\pri}, \bm{\mathcal{F}}_{-m}}_{\mr{cl}} \Exp{i (m - m\pri) \omega t}
	 \over 2 m (m - m\pri) \omega^2} \right\}
	 , \nonumber \\
	\mathcal{D}_{F, \mr{mic}}^{(2)} &= 
	- \sum_{m\ne 0} 
	{\mr{diff}\rbr{\bm{\mathcal{F}}_{-m}, \mathcal{D}} \Exp{i m \omega t}
	\over (m \omega)^2} 
	. 
\end{align}
In this case, the renormalized drift field $\bm{f}$ and the diffusion matrix $G$ are determined from Eq.~\eqref{eq: fokker, coef}.

\section{\label{sec: conv}Convergence property of the FM expansion}
Mathematically, the FM expansion of a master equation is guaranteed to converge if its Fokker-Planck operator $\mL_t$ satisfies 
\begin{align}
	\int_0^T dt \| \mathcal{L}_t \| \le \zeta
	, \label{eq: fme, 13}
\end{align}
where $\| \cdot  \|$ is the operator norm and $\zeta = \mathcal{O}(1)$ is a universal constant \cite{BlanesS09}. 
There are two problems on applying the bound \eqref{eq: fme, 13} to classical systems. 
First, $\| \mathcal{L}_t \|$ and hence the left-hand side of \eqref{eq: fme, 13} grows linearly with system size for a many-body system. 
Second, $\| \mathcal{L}_t \|$ usually contains unbounded operators like a derivative operator $\partial_{\phi_i}$. 
Although these two problems make rigorous discussions on the validity and the convergence of the FM expansion difficult, the FM expansion is found to be valid for non-chaotic few-body systems and generic many-body systems as we will see below. 

\subsection{Few-body system}
For a few-body systems, where the only second problem arises, the FM expansion is expected to be convergent for a non-chaotic system, typically for a system under a sufficiently strong dissipation. 
In this system, $\mL_t$ is expected to be a Lyapunov continuous: 
\begin{align}
	\left| \mL_t P_0 - \mL_t P_0\pri \right|
	\le C_t
	\left| P_0 - P_0\pri \right|
	, 
\end{align}
where $P_0$ and $P_0\pri$ are some probability distributions, and $C_t$ is some constant independent of $\omega$. 
Therefore, when $\omega$ is sufficiently large such that $\omega \ge (2\pi \max_{0 \le t \le T} C_t) / \zeta$, we have 
\begin{align}
	\int_0^T dt \| \mathcal{L}_t \|
	\le 
	T \max_{0 \le t \le T} C_t \le \zeta
	. \label{eq: fme, 16}
\end{align}
Thus, Eq.~\eqref{eq: fme, 13} is satisfied. 
The above discussions is consistent with the previous studies on the chaos and the bifurcation in periodically driven systems \cite{BlackburnJ92, BartuccelliM01, MannB06, ButikovE11}. 

On the other band, the FM expansion is useless for predicting the long-time behavior of a chaotic system irrespective to whether it is convergent or not. 
Let $L_{TR}^{(m_0)}$ and $P_0$ be the truncated FM expansion with truncation order $m_0$ and some initial probability distribution, respectively. 
Then, two probability distributions at time $t$ with different truncation orders $m_0\pri$ and $\widetilde{m}_0$, i.e., $\exp\br{L_{TR}^{(m_0\pri)}t}P_0$ and $\exp\br{L_{TR}^{(\widetilde{m}_0)}t}P_0$, are quite different for large $t$ due to the chaotic nature. 
This indicates that the time evolution strongly depends on the truncation order $m_0$ and therefore the FM expansion is useless.

%

\subsection{Many-body system: preliminaries and statements}
In a many-body classical system, there appear both problems, the extensiveness of $\| \mathcal{L}_t \|$ and the presence of unbounded operators. 
Although the first problem appears even in interacting quantum spin or fermionic systems, rigorous results on the energy absorption and the existence of the Floquet prethermal states are obtained by fully utilizing the boundedness of their local operators \cite{KuwaharaT16, MoriT16, AbaninD15, AbaninD17}. 
On the other hand, it is quite hard to obtain a similar bound on the classical systems because unbounded operators are notoriously difficult to handle even in mathematics. 
Nevertheless, as we will see below, we can estimate the higher-order terms in the FM expansion by combining the dimensional analysis with the techniques developed in quantum systems \cite{KuwaharaT16, MoriT16}. 
We note that such a general discussion is hard to obtain within the framework of the multi-scale perturbation theory \cite{BenderC13, StrogatzS18} because its calculation becomes involved even in the low orders. 
From this analysis, we argue that the FM expansion is, at least asymptotically, convergent. 
Moreover, its truncated series is found to well describe the steady state for a generic many-body system including a prethermal state for an isolated system and a NESS for a driven dissipative system.

Let us consider the EOM \eqref{eq: sto, eom, field} for a classical field $\bmphi_{\bmr}$. 
We assume that the drift field $\bmf_{\bmr}$ and the diffusion matrix $G_{\bmr}$ depend on the fields $\bmphi_{\bmr\pri}$ residing on, at most, $k$ neighboring sites of $\bmr$, 
which implies that the interaction and diffusion are, at most, $k$-body. 
That is, $\bmf_{\bmr}$ and $G_{\bmr}$ are functions of $\bmphi_{\bmr_1}, \bmphi_{\bmr_2}, \cdots, \bmphi_{\bmr_{k\pri}}$, and $t$ with $k\pri \le k$: 
\begin{align}
	\bmf_{\bmr} &\equiv
	\bmf_{\bmr}\br{
	\bmphi_{\bmr_1}, \bmphi_{\bmr_2}, \cdots, \bmphi_{\bmr_{k\pri}}, t}
	, \nonumber \\
	G_{\bmr} &\equiv
	G_{\bmr}\br{
	\bmphi_{\bmr_1}, \bmphi_{\bmr_2}, \cdots, \bmphi_{\bmr_{k\pri}}, t}
	. 
\end{align}
Note that this condition is automatically satisfied for finite-range interactions and diffusion. 
We introduce a dimensionless field 
$\wt\bmphi_{\bmr} := \br{
\bmphi_{\bmr, 1}/\phi_{0, 1}, 
\bmphi_{\bmr, 2}/\phi_{0, 2}, \cdots
\bmphi_{\bmr, N_I}/\phi_{0, N_I}}$, with $\phi_{0,i}$ being the typical magnitude of $\bmphi_{\bmr, i}$, and rescale the random fields $\wt\bmh_{\bmr} := \bmh_{\bmr}/\sqrt{D}$. 
For example, for the case of the GP equation \eqref{eq: FME, GP}, $\bmphi_{\bmr} = \psi_{\bmr}$ is rescaled by $\sqrt{\rho_0}$, with $\rho_0$ being the average density of a condensate. 
Then, the rescaled EOM is given by $d \wt\bmphi_{\bmr} / dt = \wt\bmf_{\bmr} + \wt G_{\bmr} \wt\bmh_{\bmr}$, 
where $\wt\bmf_{\bmr}$ and $\wt G_{\bmr}$ are the rescaled drift field and diffusion matrix, respectively. 
The corresponding master equation is given by 
\begin{align}
	{\partial P \over \partial t}
	 &= \sum_{\bmr}\rbr{
	{\partial \over \partial \wt\phi_{\bmr, a}} \br{\wt{\mathcal{F}}_{\bmr, a}P}
	 + {\partial^2 \over \partial \wt\phi_{\bmr, a} \partial \wt\phi_{\bmr, b}} 
	\br{\wt{\mathcal{D}}_{ab}P}}
	\nonumber \\
	 &=: \sum_{\bmr} \hat{L}_{\bmr}(t) P
	, \label{eq: fme, 1}
\end{align}
where the local operator $\hat{L}_{\bmr}(t)$ acts on, at most, $k$ neighboring sites of $\bmr$. 
More generally, we define the locality of the operator as follows: an operator $A := \sum_{\bmr}A_{\bmr}$ is 
said to be $k_A$-local if $A_{\bmr}$ depends on, at most, $k_A$ neighboring sites of $\bmr$:
\begin{align}
	A_{\bmr} &\equiv
	A_{\bmr}\br{
	\bmphi_{\bmr_1}, \bmphi_{\bmr_2}, \cdots, \bmphi_{\bmr_{k_A\pri}}, t}
	, 
\end{align}
where $k_A\pri$ is an integer that satisfies $k_A\pri \le k_A$. 
According to this definition, the Fokker-Planck operator $\hat{L}(t) := \sum_{\bmr} \hat{L}_{\bmr}(t)$ is a $k$-local operator. 
Since $\wt\phi_{\bmr}$ is dimensionless, $\hat{L}_{\bmr}(t)$ has the physical dimension of frequency, where we write its typical magnitude as $\omega_0$.

Formally, the time evolution operator $U(t,0)$ of Eq.~\eqref{eq: fme, 1} is given by the Dyson series: 
\begin{align}
	U(t,0) &= \sum_{m=0}^\infty {1 \over m!} 
	\int_0^tdt_1 \cdots \int_0^tdt_m
	\mathcal{T}_t \rbr{
	\hat{L}(t_1)\cdots \hat{L}(t_m)
	}
	\nonumber \\
	&= 
	\mathcal{T}_t 
	\exp\rbr{\int_0^t \hat{L}(t\pri) dt\pri}
	. 
\end{align}
Then, the exponent $\Omega(t)$ defined by $U(t,0) =: \exp\rbr{\Omega(t)}$ satisfies the following differential relation \cite{BlanesS09}: 
\begin{align}
	{d \Omega(t) \over dt} = 
	\sum_{m=0}^\infty {B_m \over m!}\mr{ad}_{\Omega(t)}^m\hat{L}(t)
	, \label{eq: fme, 2}
\end{align}
where $\mr{ad}_{\Omega}\hat{L} := \rbr{\Omega, \hat{L}}$ and $B_m$ is the $m$th Bernoulli number.

Let us formally expand the effective operator $\Omega_F:=\Omega(T)$ in powers of $\omega^{-1}$ as follows: 
\begin{align}
	\Omega_F = \sum_{m=0}^\infty \Omega_F^{(m)}
	, \label{eq: fme, 3}
\end{align}
where $\Omega_F^{(m)} = \mathcal{O}(\omega^{-m})$. 
The FM expansion of the effective Fokker-Planck operator $\hat{L}_F := \Omega_F/T$ is obtained by iteratively substituting Eq.~\eqref{eq: fme, 3} into Eq.~\eqref{eq: fme, 2} followed by the integration over $t$ from $t=0$ to $t=T$. 
We denote the $m$th-order truncated series of $\Omega_F$ as $\Omega_{TR}^{(m)}$ and the time-evolution operator generated by it as $U^{(m)}(t)$: 
\begin{align}
	\Omega_{TR}^{(m)} := \sum_{m\pri=0}^{m} \Omega_F^{(m\pri)}
	, \quad
	U^{(m)}(t) := \exp\rbr{{t\Omega_{TR}^{(m)} \over T}}
	. 
\end{align}
If this formal expansion converges, the exact time evolution $U(t,0)$ is well approximated by $U^{(m)}(t)$. 
For a generic many-body system, due to the non-integrability of the system, the system equilibrates with some typical timescales, which we denote by $\tau$. 
Then, the two steady states $P_{SS}^{(m)} := U^{(m)}(t)P_0$ and $P_{SS} := U(t,0)P_0$ with $t \gg \tau$ do not depend on an initial probability distribution $P_0$. 
We note that $\tau$ defines the timescale of the initial relaxation to the Floquet prethermalization for an isolated system and that of the relaxation to the NESS for an open system.

In what follows, we will claim the following two statements by evaluating $\Omega_F^{(m)}$ under the assumption that $\omega$ is much larger than $k \omega_0$. 
(i) The formal expansion \eqref{eq: fme, 3}, at least asymptotically, converges up to the order $m_0 \approx \omega/(k\omega_0)$: 
\begin{align}
	\left| 
	\Omega_{TR}^{(m)} - \Omega_{TR}^{(m_0)}
	\right| = 
	\mathcal{O}\rbr{\br{ k \omega_0 / \omega}^{m_0+1}}
	. \label{eq: validity, 1}
\end{align}
(ii) The exact steady state $P_{SS}$ is well approximated by the steady state $P_{SS}^{(m_0)}$ obtained from the truncated FM expansion with an exponentially small error: 
\begin{align}
	P_{SS}^{(m_0)} \simeq P_{SS}. 
    \label{eq: validity, 2}
\end{align}

\subsection{Many-body system: derivation}
The $m$th-order term $\Omega_F^{(m)}$ with $m \ge 1$ is given from the FM expansion as follows \cite{BialynickiI69}:
\begin{align}
	\Omega_F^{(m)} &= 
	\sum_{\sigma \in S_m} 
	{(-1)^{m-\Theta_\sig} \Theta_\sig!
	(m-\Theta_\sig)! \over (m+1)^2 m!}
	\nonumber \\
	&\times
	\int_0^T dt_{m+1}
	\cdots 
	\int_0^{t_2} dt_1
	\mr{ad}_{\hat{L}_{m+1}}
	\mr{ad}_{\hat{L}_m} 
	\cdots
	\mr{ad}_{\hat{L}_2}\hat{L}(t_{\sig(1)})
	, \label{eq: fme, 4}
\end{align}
where $S_m$ is the permutation group with order $m$ and 
$\hat{L}_i$ and $\Theta_\sig$ are defined by 
$\hat{L}_i := \hat{L}(t_{\sig(i)})$ and $\Theta_\sig := \sum_{i=1}^m\theta\rbr{\sig(i+1) - \sig(i)}$, with $\theta(x)$ being the Heaviside unit step function.

The typical order of the commutators in Eq.~\eqref{eq: fme, 4} is estimated from the locality of $\hat{L}_{\bmr}$ as follows. 
Let us take two operators $A = \sum_{\bmr}A_{\bmr}$ and $B = \sum_{\bmr}B_{\bmr}$, and consider their commutator $\rbr{A,B}=:\sum_{\bmr}C_{\bmr}$.  
If we assume that $A$ ($B$) is $k_A$-local ($k_B$-local) and that its typical magnitude is $g_A$ ($g_B$), 
their commutator is $(k_A + k_B)$-local and the typical magnitude of $C_{\bmr}$ is $(k_A + k_B)g_Ag_B$. 
The commutator $\rbr{\hat{L}_2, \hat{L}_1} := \mr{ad}_{\hat{L}_2}\hat{L}(t_{\sig(1)}) =:\sum_{\bmr}l_{\bmr}$ is $(2k)$-local and the typical magnitude of $l_{\bmr}$ is $2k \omega_0^2$. 
By iteratively using this, the $m$-fold commutator in Eq.~\eqref{eq: fme, 4} is $(m+1)k$-local and its typical magnitude 
is estimated as follows: 
\begin{align}
	\mr{ad}_{\hat{L}_{m+1}}
	\mr{ad}_{\hat{L}_m} 
	\cdots
	\mr{ad}_{\hat{L}_2}\hat{L}(t_{\sig(1)})
	 = \mathcal{O}\rbr{N (\omega_0 k)^{m+1} (m+1)!}
	. \label{eq: fme, 7}
\end{align}
Combining Eq.~\eqref{eq: fme, 7} with the inequality $(-1)^{m-\Theta_\sig} \Theta_\sig!
	(m-\Theta_\sig)! \le m!/2^m$, we obtain 
\begin{align}
	\left| \Omega_F^{(m)} \right| &\lesssim 
	m! \times {1 \over (m+1)^2 m!} 
	\times {m! \over 2^m} \times {T^{m+1} \over (m+1)!} 
	\nonumber \\
	&\times (\omega_0 k)^{m+1} (m+1)! N
	\nonumber \\
	&=
	{N m! \over (m+1)^2} \br{{\pi k \omega_0 \over \omega}}^{m+1}
	. \label{eq: fme, 6}
\end{align} 
This indicates that the $m$th-order term describes the collective motion of the fields $\wt\bmphi_{\bmr}$ on $(m+1)k$ sites induced by a drive and that such a process is suppressed exponentially up to the order $m = m_0 \simeq \omega/(k\omega_0)$. 
By taking $m = m_0$, we obtain $|\Omega_F^{(m)}| \lesssim N \Exp{- \zeta m}$ with constant number $\zeta = \mathcal{O}(1)$; thus we have
\begin{align}
	\left|\Omega_{TR}^{(m)}- \Omega_{TR}^{(m_0)}\right|
	 = 
	\mathcal{O}\br{ N \Exp{- \zeta m} }
	, \label{eq: fme, 9}
\end{align}
which indicates that the truncated series $\Omega_{TR}^{(m)}$ seems to converge up to the order $m \le m_0$. 
Thus, we have complete the derivation of Eq.~\eqref{eq: validity, 1}. 

Next, we evaluate the difference between $U^{-1}(T,0)\Omega_{TR}^{(m_0)}U(T,0)$ and $\Omega_{TR}^{(m_0)}$. 
Let us expand them in powers of $T$ as follows: 
\begin{align}
	&U^{-1}(T,0)\Omega_{TR}^{(m_0)}U(T,0)
	\nonumber \\
	 &= \sum_{m=0}^\infty 
	{1 \over m!} 
	\int_0^Tdt_1 \cdots \int_0^Tdt_m
	\mathcal{T}_t \rbr{
	\mr{ad}_{L(t_1)}\cdots \mr{ad}_{L(t_m)}
	\Omega_{TR}^{(m_0)}
	}
	\nonumber \\ 
	&=:
	\sum_{m=0}^\infty \mathcal{A}_m\Omega_{TR}^{(m_0)}
	, 
	\nonumber \\
	&\Omega_{TR}^{(m_0)} = 
	\rbr{U^{(m_0)}(T)}^{-1}\Omega_{TR}^{(m_0)}U^{(m_0)}(T)
	\nonumber \\
	 &= \sum_{m=0}^\infty 
	\sum_{r=0}^m
	\sum_{ \tbr{l_i}_{i=1}^r}
	{1 \over r!}
	\mr{ad}_{\Omega_F^{(l_1)}}\cdots 
	\mr{ad}_{\Omega_F^{(l_r)}}
	\Omega_{TR}^{(m_0)}
	\nonumber \\
	&=: 
	\sum_{m=0}^\infty \mathcal{A}\pri_m\Omega_{TR}^{(m_0)}
	, 
\end{align}
where $\sum_{ \tbr{l_i}_{i=1}^r}$ denotes the sum over all the sets of integers $\tbr{l_i}_{i=1}^r$ that satisfy $\sum_{i=1}^r(l_i+1) = m$ and $0\le l_i \le m_0$. 
In the iterative procedure in determining $\Omega_F^{(m_0)}$ in the FM expansion, $\Omega_F^{(m_0)}$ is chosen such that $\mathcal{A}\pri_m\Omega_{TR}^{(m_0)}$ coincides with $\mathcal{A}_m\Omega_{TR}^{(m_0)}$ for any $m \le m_0$; thus we have 
\begin{align}
	&U^{-1}(T,0)\Omega_{TR}^{(m_0)}U(T,0) - 
	\Omega_{TR}^{(m_0)}
	\nonumber \\
	&= \sum_{m=m_0+1}^\infty
	\br{\mathcal{A}_m\Omega_{TR}^{(m_0)} - \mathcal{A}\pri_m\Omega_{TR}^{(m_0)}}
	. \label{eq: fme, 5}
\end{align}
Using the argument for folded commutators around Eq.~(\ref{eq: fme, 7}), we obtain 
\begin{align}
	\left|\mathcal{A}_m\Omega_{TR}^{(m_0)}\right|
	 &\lesssim
	\br{{4 \pi k \omega_0 \over \omega}}^m 
	\left|\Omega_{TR}^{(m_0)} \right|
	, \nonumber \\
	\left|\mathcal{A}\pri_m\Omega_{TR}^{(m_0)}\right|
	 &\lesssim 
	\br{{8 \pi k \omega_0 \over \omega}}^m
	\left|\Omega_{TR}^{(m_0)}\right|
	. 
\end{align}
Thus, we have 
\begin{align}
	&\left|U^{-1}(T,0)\Omega_{TR}^{(m_0)}U(T,0) - 
	\Omega_{TR}^{(m_0)} \right|
	\nonumber \\
	 &\lesssim 
	\br{{8 \pi k \omega_0 \over \omega}}^{m_0} 
	\left|\Omega_{TR}^{(m_0)}\right|
	 \lesssim 
	N \Exp{- \zeta m_0}
	, 
\end{align}
where we used Eq.~\eqref{eq: fme, 6} in the last relation. 
Finally, we obtain 
\begin{align}
	&\left|U^{-1}(n_t T, 0)\Omega_{TR}^{(m_0)}U(n_t T,0) - 
	\Omega_{TR}^{(m_0)}\right|
	\nonumber \\
	 &\lesssim 
	n_t \left|U^{-1}(T,0)\Omega_{TR}^{(m_0)}U(T,0) - 
	\Omega_{TR}^{(m_0)}\right|
	\lesssim
	N n_t \Exp{- \zeta m_0}
	, \label{eq: fme, 8}
\end{align}
where $n_t$ is an integer and $\zeta$ is a constant number. 
Consider two autonomous equations
\begin{align}
	{d P(s) \over ds} &= 
	\Omega_{TR}^{(m_0)}P(s)
	, \nonumber \\
	{d P(s) \over ds} &= 
	U^{-1}(n_t T, 0)\Omega_{TR}^{(m_0)}U(n_t T,0)
	P(s)
	, \label{eq: fme, 11}
\end{align}
with the same initial probability distribution $P_0$: $P(s=0) = P_0$, where $s$ is an auxiliary time. 
For a sufficiently large frequency $\omega$, the relaxation timescale $\tau$ satisfies 
\begin{align}
	\omega \tau 
	\ll \Exp{\zeta m_0} 
	\simeq \Exp{{\zeta \omega \over k \omega_0}}
	. 
\end{align}
Then, from the bound \eqref{eq: fme, 8}, the solutions $U^{(m_0)}(s)P_0$ and $U^{-1}(n_t T, 0)U^{(m_0)}(s)U(n_t T,0)P_0$ of Eq.~\eqref{eq: fme, 11} at time $s \gtrsim \tau$ shows almost similar thermodynamics properties: 
\begin{align}
	U^{(m_0)}(s)P_0 \simeq 
	U^{-1}(n_t T, 0)U^{(m_0)}(s)U(n_t T,0)P_0
	. \label{eq: fme, 12}
\end{align}
By applying $U(n_t T, 0)$ to the both sides of Eq.~\eqref{eq: fme, 12} from the left, we have 
\begin{align}
	U(n_t T, 0)U^{(m_0)}(s)P_0 \simeq 
	U^{(m_0)}(s)U(n_t T,0)P_0
	. \label{eq: fme, 14}
\end{align}
For a generic many-body system, the state after the relaxation $n_t T \gtrsim \tau$ ($s \gtrsim \tau$), $U(n_t T, 0)U^{(m_0)}(s)P_0$ ($U^{(m_0)}(s)U(n_t T,0)P_0$) approaches the steady state $P_{SS}$ ($P_{SS}^{(m_0)}$) that is solely determined from $U(n_t T, 0)$ ($U^{(m_0)}(s)$): 
\begin{align}
	U(n_t T,0)U^{(m_0)}(n_tT)P_0 &= P_{SS}
	, \nonumber \\
	U^{(m_0)}(n_tT)U(n_t T,0)P_0 &= P_{SS}^{(m_0)}
	. \label{eq: fme, 15}
\end{align}
Combining Eqs.~\eqref{eq: fme, 14} and \eqref{eq: fme, 15}, we obtain $P_{SS} = P_{SS}^{(m_0)}$, which completes the derivation of Eq.~\eqref{eq: validity, 2}.

\subsection{Many-body system: discussion}
For an isolated Hamilton system, it is clear from Eq.~\eqref{eq: fme, 8} that the truncated FM Hamiltonian $H_{TR}^{(m_0)} := i \Omega_{TR}^{(m_0)}/T$ is a quasi-conserved quantity, where the transient state before heating is given by the Gibbs distribution 
$P_{SS} \propto \exp(- H_{TR}^{(m_0)}/T_{\rm te})$ \cite{MoriT18}. 
From Eq.~\eqref{eq: fme, 9}, $H_{TR}^{(m_0)}$ and hence the prethermal state are well approximated by the lower-order truncation $H_{TR}^{(m)}$ and the Gibbs distribution of $H_{TR}^{(m)}$, respectively. 
We note that while its macroscopic properties are well captured by the truncated FM expansion, its local dynamics is not, in contrast to quantum systems due to the onset of chaos \cite{MoriT18}. 

Next, we consider an isolated system that is not a Hamilton system, such as a general dynamical system \cite{WigginsS03} and a stochastic process, e.g., the asymmetric simple exclusion process (ASEP) \cite{LiggettT12, SpohnH12}. 
For these systems, there is no \textit{a priori} method to determine their steady state like the equipartition principle in Hamilton systems. 
Nevertheless, the above results tell us that the exact steady states are well approximated from the truncated FM expansion $\Omega_{TR}^{(m_0)}$. 
From Eq.~\eqref{eq: fme, 9}, $\Omega_{TR}^{(m_0)}$ is well approximated by the lower-order truncation $\Omega_{TR}^{(m)}$ and hence we expect that the exact steady states are well approximated by $\Omega_{TR}^{(m)}$. 
If the FM expansion is divergent, the system is expected to approach a featureless state with a chaotic nature which is an analog of an infinite-temperature state in an isolated Hamilton system. 
The above discussion implies that a transient state described by the truncated series $\Omega_{TR}^{(m)}$ can appear before approaching the featureless state, which is reminiscent of a Floquet prethermal state in an isolated Hamilton system.

Finally, we discuss a driven dissipative system. 
Due to the balance between the drive and the damping, the system relaxes into the NESS with time $\tau$. 
The steady state $P_{SS}$ is, therefore, not a transient state but a NESS and 
the truncated FM expansion well approximates $P_{SS}$ in the high-frequency regime.
This fact implies that the truncated FM expansion well describes the overall dynamics of the system from $t=0$ to $t=\infty$. 
We note that, though the above derivation is conducted with classical systems in mind, similar results are obtained for open quantum systems by replacing a (classical) master equation with a quantum master equation \cite{BreuerH02}. 
Therefore, the FM expansion is useful even in open many-body systems, although the validity of the FM expansion for the quantum master equation is confirmed only in few-body systems \cite{HaddadfarshiF15, RestrepoS16, DaiC16, DaiC17, SchnellA18}.


\section{\label{sec: Kapitza}Kapitza pendulum with friction}
In this section, we take the Kapitza pendulum with friction as an example of a driven dissipative few-body system to test the validity of our FM expansion as its effective description. 

\subsection{Setup}
\begin{figure}[t]
\centering
\includegraphics[width=\columnwidth, clip]{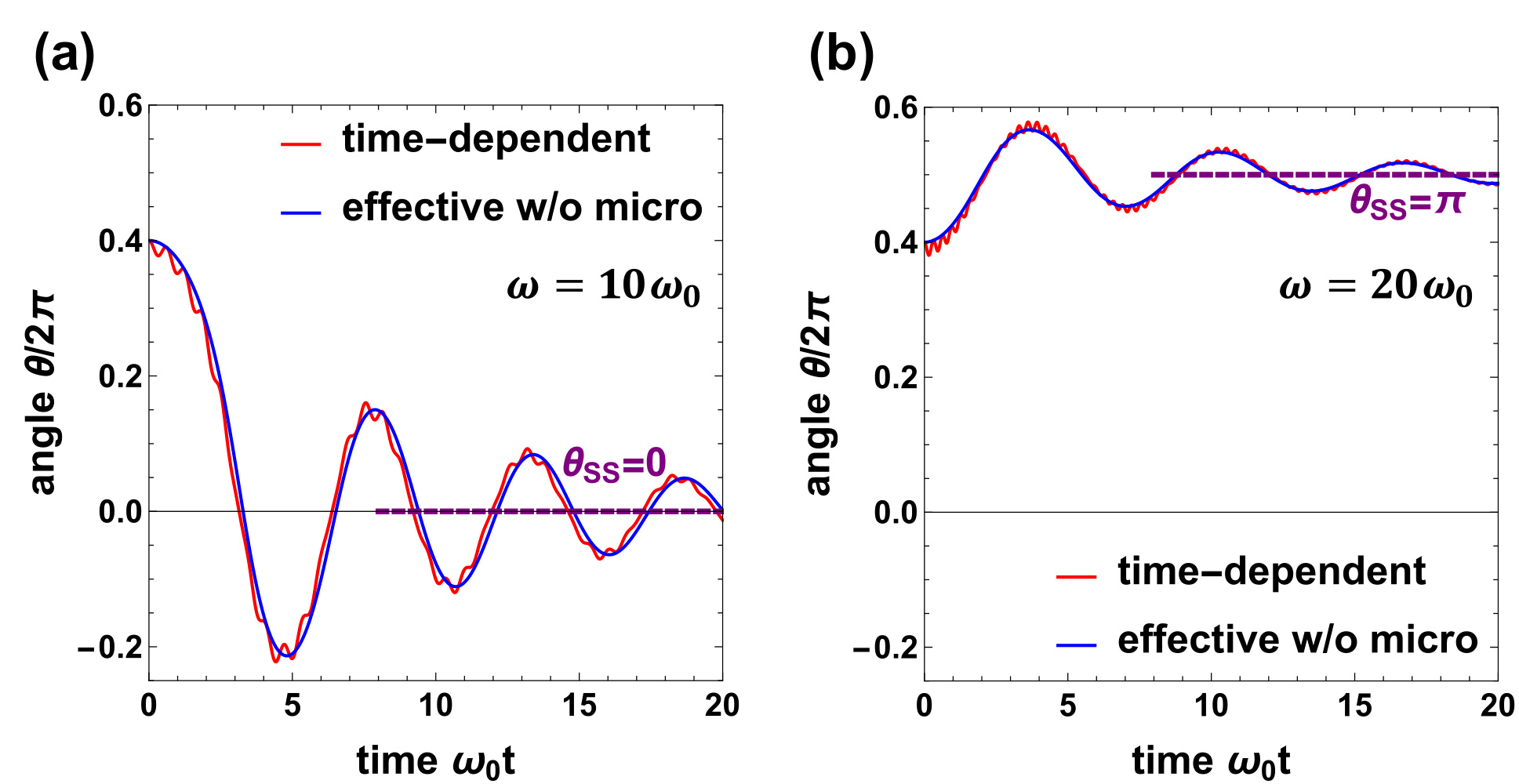}
\caption{
Time evolution of angle $\theta(t)$ for (a) slow ($\omega/\omega_0 = 10$) and (b) fast ($\omega/\omega_0 = 20$) drive, where the parameters are chosen as $a/l = 0.1$ and $\gamma/\omega_0 = 0.2$. 
	The red and blue curves are obtained by solving the time-dependent EOM \eqref{eq: kap, damp} and the effective EOM \eqref{eq: kap, damp, eff}, respectively, which are in excellent agreement. 
	Starting from the same initial state $\theta_0 = 0.8 \pi$ with $v = 0$, the angle approaches $0$ for the slow drive (a), while it approaches $\pi$ for the fast drive (b), as shown in the purple lines.  
	The purple dashed lines show the steady-state angles $\theta_{SS}$, which is either $0$ or $\pi$. 
	}
       \label{fig: Kapitza}
\end{figure}

The Kapitza pendulum \cite{KapitzaP51} is a classical rigid pendulum with a vertically oscillating point of suspension (see Fig.~\ref{fig: Kapitza, setup}), where $\theta$ is the angle measured from the downward position, $\omega_0 = \sqrt{g / l}$ is the frequency of the small oscillations near $\theta = 0$ ($g$ and $l$ are the gravitational constant and the length of the pendulum, respectively).
The suspension point oscillates with amplitude $a$ and frequency $\omega$: $y_c = -a \cos(\omega t)$. 
Its EOM reads \cite{KapitzaP51, DAlessioL13, BukovM15}
\begin{align}
	\ddot{\theta} =
	 - \rbr{\omega_0^2 + {a \over l} \omega^2 \cos(\omega t)}
	 \sin\theta
	. 
	\label{eq: kap, original}
\end{align}
While the first term on the right-hand side describes the gravitational force, the second one comes from the inertial force of the oscillation of the suspension point. 
As is first shown by Kapitza, the highest point $\theta = \pi$ becomes stable above the critical frequency $\omega_c = \sqrt{2} l \omega_0 / a$, and the pendulum performs oscillations around this inverted position. 
From the Floquet-engineering viewpoint, the dynamics of the pendulum is described by a time-independent effective Hamiltonian, with its effective potential developing a local minimum at $\theta = \pi$ for $\omega > \omega_c$ \cite{KapitzaP51, DAlessioL13, BukovM15}. 
The Kapitza pendulum is a prototypical example of dynamical stabilization, a stabilization of a system by a periodic drive, which is widely employed in many areas of physics \cite{CourantE52, FeynmanR64, PaulW90, CasatiG94, WiedemannH94, ReinholdC97} including beam focusing in a synchrotron (alternating-gradient focusing \cite{CourantE52, FeynmanR64}), and trapping ions in the Paul trap \cite{PaulW90}. 

We introduce a friction term $-\gamma \dot{\theta}$ to make sure that the system reaches to a stable point after a long time. The EOM with friction is given by
\begin{align}
	\ddot{\theta} = - \gamma \dot{\theta}
	 - \rbr{\omega_0^2 + {a \over l} \omega^2 \cos(\omega t)}\sin\theta
	. 
	\label{eq: kap, damp, pre}
\end{align}
We note that Eq.~\eqref{eq: kap, damp, pre} is no longer written in a Hamilton equation due to the friction term $-\gamma \dot{\theta}$. 
We note that this model is previously studied in the context of the chaos and the bifurcation theory \cite{BlackburnJ92, BartuccelliM01, MannB06, ButikovE11}. 
We revisit this problem from the viewpoint of Floquet engineering and test the validity of the FM expansion. 
In the following, we analyze Eq.~\eqref{eq: kap, damp, pre} using the FM expansion in the previous section and confirm that it correctly reproduces the time evolution of the pendulum and the stability at the inverted point $\theta = \pi$. 

\begin{figure}[t]
\centering
\includegraphics[width=0.5\columnwidth, clip]{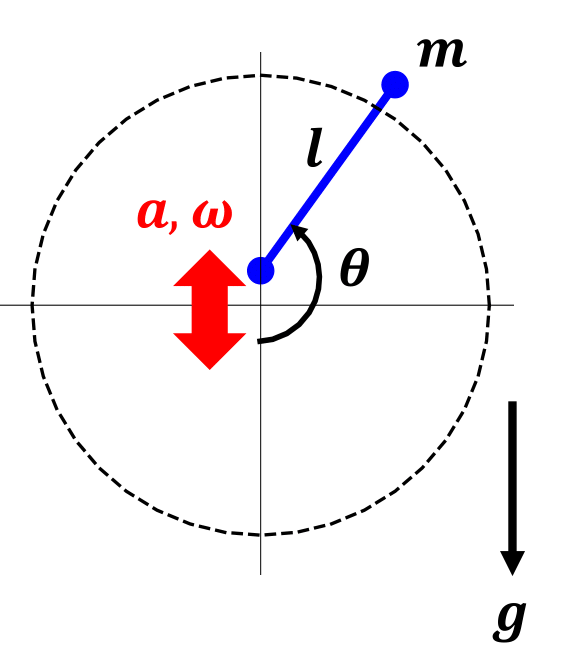}
\caption{
Schematic illustration of the Kapitza pendulum, where $a$ and $\omega$ are the amplitude and frequency of the vertical oscillation of the suspension point, respectively. 
Here, $\theta, l$, and $g$ are the angle measured from the downward position, the length of the pendulum, and the gravitational constant, respectively. 
}
       \label{fig: Kapitza, setup}
\end{figure}  

\subsection{FM expansion and effective EOM}
To apply the general formalism developed in Sec.~\ref{sec: FME, cla, det}, we rewrite Eq.~\eqref{eq: kap, damp, pre} into a first-order ordinary differential equation of $\theta$ and $v$ ($v$ is the angular velocity) as follows: 
\begin{align}
	\dot{\theta} &= v
	\nonumber \\
	\dot{v} &= 
	- \gamma v
	 - \rbr{\omega_0^2 + {a \over l} \omega^2 \cos(\omega t)} \sin\theta
	. 
	\label{eq: kap, damp}
\end{align}
Comparing with Eq.~\eqref{eq: sto, eom}, we find that the classical variable $\bmphi$ in Eq.~\eqref{eq: kap, damp} consists of a two-dimensional vector: $\bmphi = (\theta, v)$. 
The Fourier components of the drift force $\bm{f}(\bmphi, t) := \bm{f}_0 + \bm{f}_1 \Exp{-i\omega t} + \bm{f}_{-1} \Exp{i\omega t}$ are given by 
\begin{align}
	\bm{f}_0(v,\theta) & := 
	\twvec{\bm{f}_{0,\theta}}{\bm{f}_{0,v}}
	 = 
	\twvec{v}{- \gamma v -\omega_0^2 \sin\theta}
	, \nonumber \\
	\bm{f}_{\pm 1}(v,\theta) & := 
	\twvec{\bm{f}_{\pm 1,\theta}}{\bm{f}_{\pm 1,v}}
	 = \twvec{0}{- {a \omega^2 \over 2l} \sin \theta}
	, 
\end{align}
and the diffusion matrix $g$ vanishes. 
From the FM expansion \eqref{eq: FME, deterministic}, 
we obtain the effective drift field $\bm{f}_F^{(2)}(v,\theta)$: 
\begin{align}
	\rbr{\bm{f}_0, \bm{f}_1}_{\mr{cl}} &= 
	f_{0,\theta} {\partial \bm{f}_1 \over \partial \theta} 
	+ f_{0,v} {\partial \bm{f}_1 \over \partial v}
	- f_{1,\theta} {\partial \bm{f}_0 \over \partial \theta}
	- f_{1,v} {\partial \bm{f}_0 \over \partial v}
	\nonumber \\
	 &= {a \omega^2 \over 2l} \twvec{\sin\theta}{-v \cos\theta - \gamma \sin\theta}
	 , 
	 \nonumber \\
	\bm{f}_F^{(2)}(v,\theta) &:= 
	- {
	\rbr{\bm{f}_{-1}, \rbr{\bm{f}_0, \bm{f}_1}_{\mr{cl}} }_{\mr{cl}} + 
	\rbr{\bm{f}_{1}, \rbr{\bm{f}_0, \bm{f}_{-1}}_{\mr{cl}}}_{\mr{cl}}
	\over 2 \omega^2}
	\nonumber \\
	&= \twvec{0}{
	- \br{{a \omega \over 2 l }}^2 \sin(2\theta)
	}
	, 
\end{align}
where the effective EOM is given as follows: 
\begin{align}
	\dot{\theta} &= v
	\nonumber \\
	\dot{v} &= 
	- \gamma v - \omega_0^2 \sin\theta 
	- \br{{a \omega \over 2 l }}^2 \sin(2\theta)
	. 
	\label{eq: kap, damp, eff}
\end{align}
Comparing Eqs.~\eqref{eq: kap, damp} and \eqref{eq: kap, damp, eff}, we find that the original static potential $-\omega_0^2\cos\theta$ is replaced by the effective potential 
\begin{align}
	V_F(\theta) = 
	- \omega_0^2 \cos\theta
	- \br{{a \omega \over 2 l }}^2 \sin^2\theta
	\label{eq: kap, pendulum}
\end{align}
due to the periodic drive. 
We note that $V_F(\theta)$ is independent of the friction strength $\gamma$ and the same as the one obtained from the analysis without friction \cite{KapitzaP51, DAlessioL13}. 
Due to the second term, the effective potential develops a new local minimum at $\theta = \pi$ above critical driving frequency $\omega_c = (\sqrt{2}l \omega_0) / a$. 
Since $(\theta, v) = (0,0)$ and $(\pi,0)$ are both stationary solutions of Eq.~\eqref{eq: kap, damp, eff}, the system converges to either of these points after sufficiently long time with the help of the friction $-\gamma v$.   
The steady-state angle $\theta_{SS} := \theta(t\to \infty)$, in general, depends on $\theta_0$, $\omega$, $v(t=0)$, and $\gamma$, as we will see below.

\subsection{Comparison between the time-periodic and effective EOMs}
In what follows, we compare Eqs.~\eqref{eq: kap, damp} and \eqref{eq: kap, damp, eff} through the dynamics of $\theta(t)$ and the steady-state angle $\theta_{SS}$. 
The parameters $a/l$ and $\gamma$ are fixed as $a/l = 0.1$ and $\gamma = 0.2 \omega_0$. 
The time evolutions of $\theta(t)$ for slow ($\omega/ \omega_0 = 10$) and fast ($\omega/ \omega_0 = 20$) drives are shown in Figs.~\ref{fig: Kapitza} (a) and (b), respectively. 
The initial states are taken as $(\theta, v) = (0.8 \pi, 0)$ in both cases. 
The red and blue curves are obtained from the time-dependent EOM \eqref{eq: kap, damp} and the effective EOM \eqref{eq: kap, damp, eff}, respectively, which are in good agreement. 
After a sufficiently long time, the pendulum approaches the lowest point $\theta = 0$ for the slow drive (a) below the critical frequency $\omega_c$, while it approaches the inverted point $\theta = \pi$ for the fast drive (b). 

In Fig.~\ref{fig: Kapitza, phase} (a), we present the steady-state ``phase diagram" of the pendulum for fixed parameters $v(t=0) = 0$ and $\gamma = 0.2 \omega_0$.
We see that the inverted point $\theta = \pi$ is preferred for an initial angle close to $\pi$ with a fast drive  [shaded region in Fig.~\ref{fig: Kapitza, phase} (a)], while it approaches the lowest point $\theta = 0$ for the other parameter region. 
The boundary curve between $\theta_{SS} = 0$ and $\pi$ for the effective EOM is determined from the effective potential \eqref{eq: kap, pendulum} as follows: 
\begin{align}
	\omega > \omega_c
	 \quad \mr{and} \quad
	|\theta - \pi| < \arccos\rbr{\br{ {\omega_c \over \omega}}^2}
	. \label{eq: kap, boundary, fit}
\end{align}

Although the effective EOM \eqref{eq: kap, damp, eff} [blue line in Fig.~\ref{fig: Kapitza, phase} (a)] gives the boundary \eqref{eq: kap, boundary, fit} close to the curve obtained from the time-dependent EOM \eqref{eq: kap, damp} 
[red line in Fig.~\ref{fig: Kapitza, phase} (a)], a slight deviation can be seen even in high frequency, which might indicate the failure of the FM expansion. 
In particular, in the high-frequency limit, $\theta_{SS}/(2\pi)$ converges to $1/4$ for the effective EOM (dashed black line in Fig.~\ref{fig: Kapitza, phase} (b)), while it converges to $0.267$ for the EOM (dashed purple line in Fig.~\ref{fig: Kapitza, phase} (b)). 
This deviation is attributed to the $\omega$-dependent coefficient in the drive: $\bm{f}_{\pm 1} \propto \omega^2$. 
Since we here fix $a/l$ rather than the coefficient $a \omega^2 / (2l)$ of $\bm{f}_{\pm 1}$, the resulting potential, the second term on the right-hand-side of Eq.~\eqref{eq: kap, pendulum}, is proportional to $\omega^2$. 
However, when we take into account the kick operator $\mG_F(t)$ in Eq.~\eqref{eq: time, evo, Floquet}, we obtain excellent agreement (almost overlapped) between the EOM and the effective EOM, shown as the red and green lines in Fig.~\ref{fig: Kapitza, phase} (b), respectively. 
The above results supports the arguments in Sec.~\ref{sec: conv} that the FM expansion for a non-chaotic few-body system well approximates the dynamics up to its steady states.

\begin{figure}[t]
\centering
\includegraphics[width=\columnwidth, clip]{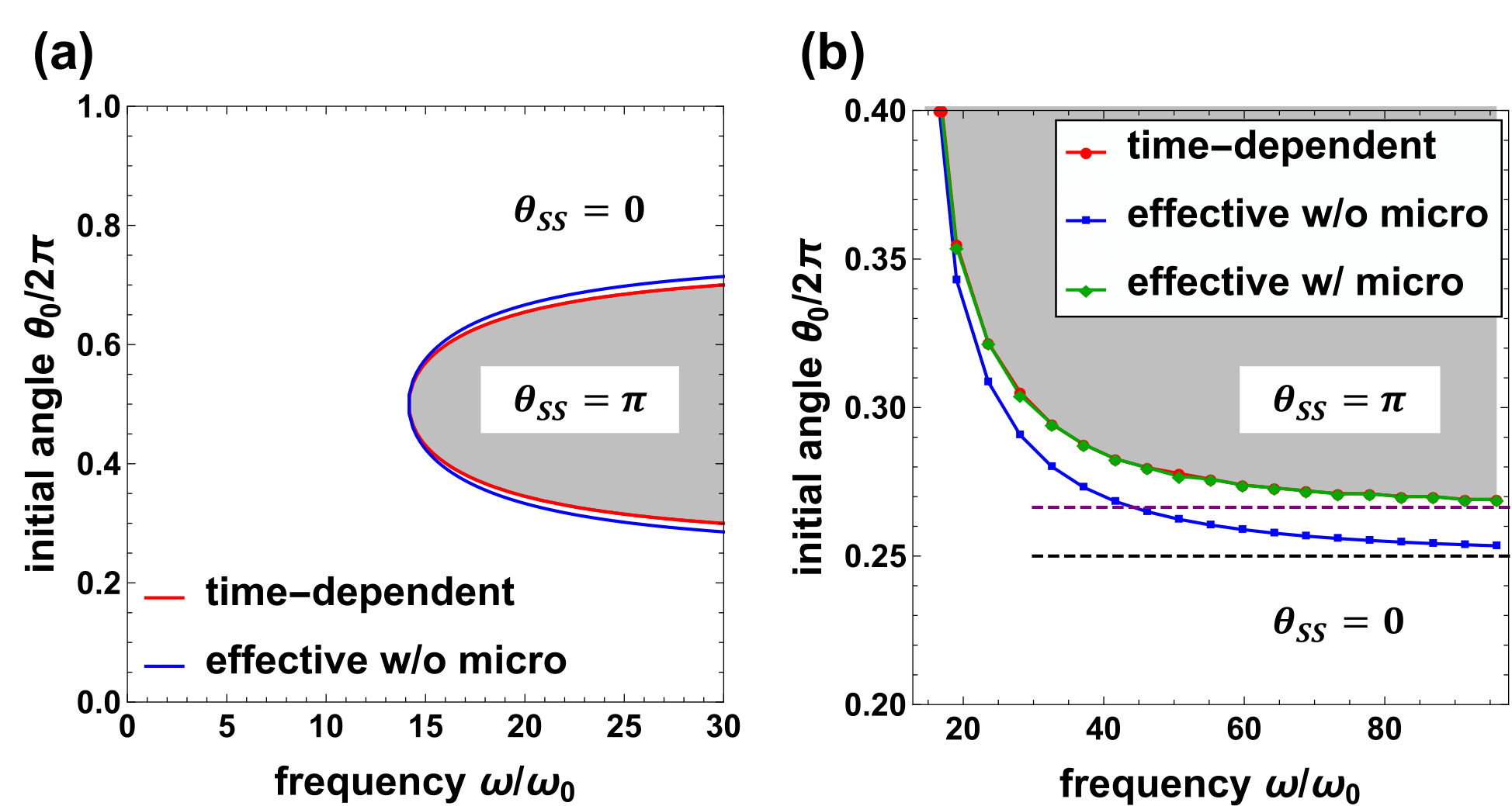}
\caption{
(a) Dependence of the steady-state angle $\theta_{SS}$ on the initial state $\rbr{\theta(0), v(0)} = (\theta_0,0)$ and frequency $\omega$, where $a/l$ and $\gamma/\omega_0$ are fixed in the same parameters as Fig.~\ref{fig: Kapitza}. 
	In the shaded region, the angle approaches $\theta = \pi$ after a sufficiently long time, while it approaches $\theta = 0$ in the other area. 
	The red and blue curves show the boundary obtained from the time-dependent EOM \eqref{eq: kap, damp} and that derived from the effective EOM \eqref{eq: kap, damp, eff}, respectively. 
(b) Boundary of the steady-state angle $\theta_{SS} = 0,\pi$ obtained from the time-dependent EOM \eqref{eq: kap, damp} (red) and the time-independent effective EOM \eqref{eq: kap, damp, eff} (blue), and the time-independent effective EOM \eqref{eq: kap, damp, eff} in addition to the kick operator (green). The red and green curves overlap almost completely. The black and purple lines are guides to the eyes. 
	}
       \label{fig: Kapitza, phase}
\end{figure}

\section{\label{sec: sLLG}Stochastic Landau-Lifshitz-Gilbert equation}
In this section, we treat a classical driven many-spin system described by the time-dependent sLLG equation 
as an example of driven many-body systems. 
We show that a circularly polarized electromagnetic wave generates the magnetization parallel to its propagating direction in classical ordered magnets, which is a manifestation of Floquet engineering of magnetizations by a laser.
We compare the time-dependent sLLG equation and the effective one obtained from the FM expansion through the time evolution of the magnetization and its long-time average at the NESS.
Through a detailed comparison changing various parameters, e.g., the frequency, the dissipation strength, and temperature, 
we find that the effective one correctly reproduces the exact time evolution for a long time until the NESS. 

While driven quantum spin systems have been studied in the context of Floquet engineering in recent years \cite{TakayoshiS14a, TakayoshiS14b, SatoM16, ClaassenM17, KitamuraS17, MentinkJ15, OkaT18}, it is theoretically hard to include the effects of dissipation and temperature in such driven quantum many-body systems. 
As we will see below, the framework based on the sLLG equation and the FM expansion enables us to treat all these effects with considerably larger system size  than quantum cases.

\begin{figure}[t!]
\centering
   \includegraphics[width=\columnwidth, clip]{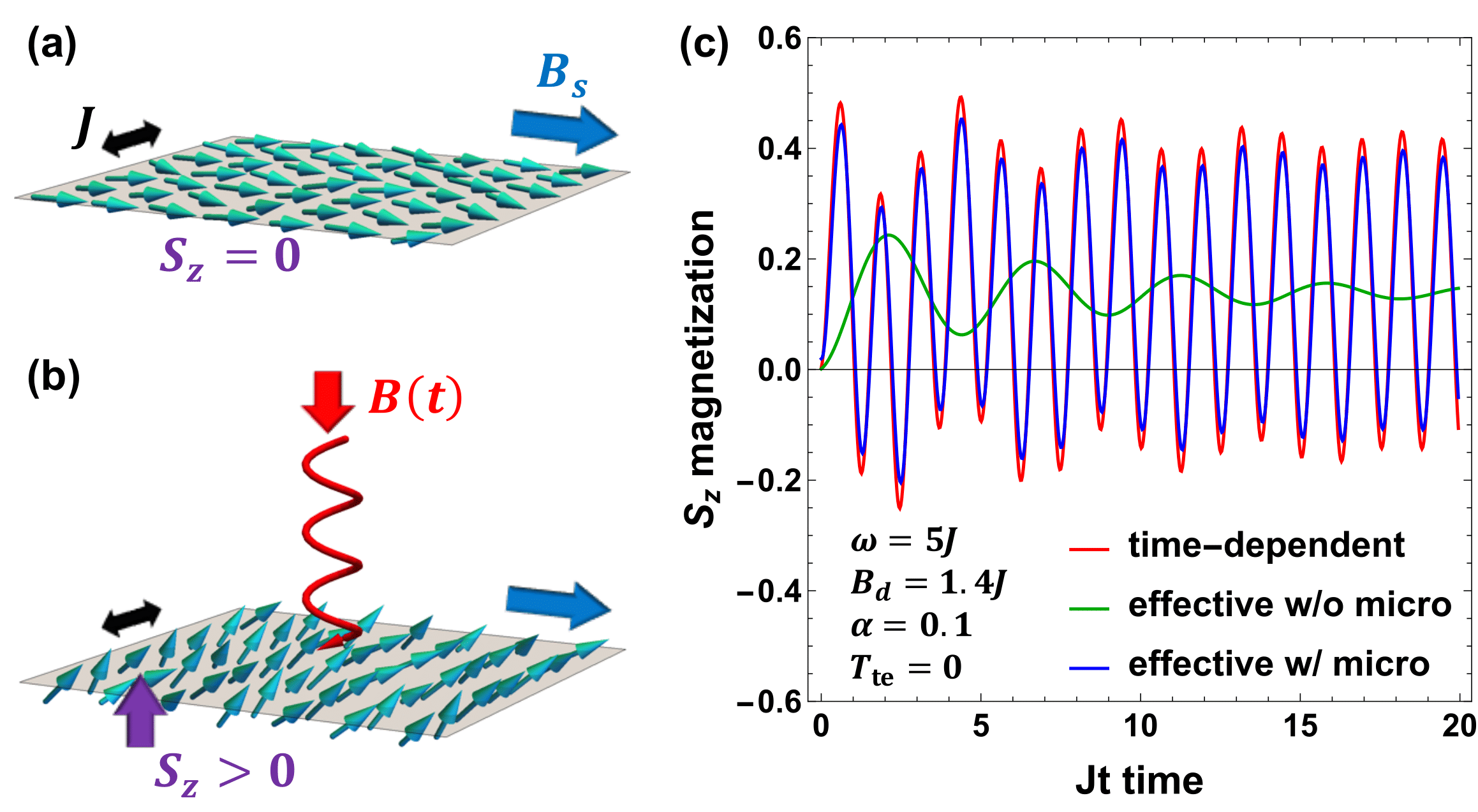}
\caption{
(a) Schematic illustration of a two-dimensional ferromagnet with $100 \times 100$ spins. 
The nearest-neighbor spins are coupled via a ferromagnetic interaction with exchange coupling $J$ and a static magnetic field with strength $B_s$ is applied in the $m_x$ direction. 
(b) When the system is irradiated by a circularly polarized magnetic field with strength $B_d$, the average magnetization $S_z$ emerges. 
(c) Time evolution of the averaged magnetization $\langle S_z \rangle$ for the original sLLG equation \eqref{eq: sLLG} (red), the effective sLLG equation \eqref{eq: eff, sLLG, 2nd} with (blue) and without (green) the kick operator. 
	The parameters are chosen as $J = 1, \omega = 5 J, B_d = 1.4 J, B_s = 1.4 J, \alpha  = 0.1$, and $T_{\mr{te}} = 0$. 
    }
       \label{fig: sLLG_setup}
\end{figure}  
\subsection{Setup}
The sLLG equation is a phenomenological equation of ferromagnets, which is formally the torque equation with a damping and a thermal fluctuation. 
It is widely employed in the field of spintronics and proved to be a powerful approach to modeling ultrafast magnetization processes like the laser-induced demagnetization of ferromagnets \cite{KirilyukA10, BeaurepaireE96, AtxitiaU09, KazantsevaN08}. 
Let $\bmm_{\bmr}$ and $\mathcal{H}(t)$ be the magnetic moment at site $\bmr$ and the Hamiltonian (energy) of the system, respectively. 
The sLLG equation reads 
\begin{align}
	\dot{\bmm}_{\bmr} =
	 - \gamma \bmm_{\bmr} \times
	\rbr{
	\bm{H}_{\bmr}(t) + \bmh_{\bmr}(t)
	}
	 + {\alpha \over m_s}
	 \bmm_{\bmr} \times \dot{\bmm}_{\bmr}
	. \label{eq: sLLG, original}
\end{align}
where $\bm{H}_{\bmr}(t) = - (\delta \mathcal{H}(t))/ (\delta \bmm_{\bmr})$ is an effective magnetic field generated by the surrounding spins and external fields, and 
$\bmh_{\bmr}(t)$ is the random magnetic field at $\bmr$ modeling the thermal fluctuation. 
Here, $\alpha, \gamma$, and $m_s := |\bmm_{\bmr}|$ are the Gilbert damping constant, the gyromagnetic ratio, 
and the magnitude of the magnetization, respectively. 
In what follows, we set $\gamma = m_s = 1$.  
The first term on the right-hand side of Eq.~\eqref{eq: sLLG, original} describes the precession around $\bm{H}_{\bmr}(t) + \bmh_{\bmr}(t)$, while the second term, which is known as the Gilbert term, describes damping toward the effective magnetic field. 
The thermal fluctuation with temperature $T_{\mr{te}}$ is modeled by $\bmh_{\bmr}(t) := \rbr{h_{\bmr, 1}(t), h_{\bmr, 2}(t), h_{\bmr, 3}(t)}$ that satisfies
\begin{align}
	\langle
	h_{\bmr, a}(t)h_{\bmr\pri, b}(t\pri)
	\rangle
	 = 
	2D \delta_{ab}
	 \delta_{\bmr, \bmr\pri}
	 \delta(t - t\pri)
	 , 
\end{align}
where $D = 2 k_B T_{\mr{te}} \alpha$ is the diffusion constant satisfying the fluctuation-dissipation theorem. 
To apply the general formalism developed in Sec.~\ref{sec: FME, cla}, we rewrite Eq.~\eqref{eq: sLLG, original} as 
\begin{align}
	\dot{\bmm}_{\bmr} &= 
	 - { \bmm_{\bmr} \over 1 + \alpha^2}
	 \times 
	\{
	\bm{H}_{\bmr}(t) + \bmh_{\bmr}(t)
	\nonumber \\
	& 
	 \quad \quad + {\alpha \over m_s}
	\bmm_{\bmr} \times 
	\rbr{ \bm{H}_{\bmr}(t) + \bmh_{\bmr}(t)}
	\}
	, \label{eq: sLLG}
\end{align}
and henceforth use this equation. 

Comparing Eqs.~\eqref{eq: sto, eom, field} and \eqref{eq: sLLG}, we find that $\bmphi_{\bmr} = \bmm_{\bmr}$ represents the spin configuration, and $\bmf_{\bmr}$ and $\bm{g}_{\bmr}$ are given by
\begin{align}
	\bmf_{\bmr}(t) & = - {\bmm_{\bmr} \over 1 + \alpha^2 } 
	\times 
	\rbr{ \bm{H}_{\bmr}(t) + {\alpha \over m_s} \bmm_{\bmr} \times \bm{H}_{\bmr}(t)}
	\nonumber \\
	&\quad + {2D \over 1 + \alpha^2} \bmm_{\bmr}
	, \nonumber \\
	g_{\bmr, ab} & = {1 \over 1 + \alpha^2} \epsilon_{abc} m_{\bmr, c} +{\alpha m_s \over 1 + \alpha^2 } \br{
	\delta_{ab} - {m_{\bmr, a} m_{\bmr, b} \over (m_s)^2}
	}
	. \label{eq: sllg, fg}
\end{align}
Here, $\bmf_{\bmr}$ and $\bm{g}_{\bmr}$ describe the spin precession generated by $\bm{H}_{\bmr}$ and the spin diffusion induced by $\bmh_{\bmr}$ and perpendicular to $\bmm_{\bmr}$, respectively \cite{LiZ04, AronC14}. 
We note that the second term in $\bmf_{\bmr}(t)$ comes from $d_k := -D g_{kl} \partial_k g_{il}$ in Eq.~\eqref{eq: fokker, coef}. 
The probability current $\bm J$ in the Fokker-Planck equation of the sLLG equation satisfies $\bm{J}\cdot\bm{m} = 0$, implying the conservation of the magnitude of the magnetic moment $d\br{\bm{m}_i, \cdot \bm{m}_i}/(dt) = 0$.

As a simple example, we here consider a classical ferromagnetic Heisenberg model on a square lattice 
[see Fig.~\ref{fig: sLLG_setup} (a)], whose Hamiltonian $\mathcal{H}(t)$ reads
\begin{align}
	\mathcal{H}(t) = 
	- J \sum_{\langle \bmr, \bmr\pri \rangle} 
	\bmm_{\bmr} \cdot \bmm_{\bmr\pri}
	- g \mu_B \sum_{\bmr} \bm{B}(t) \cdot \bmm_{\bmr} 
	, \label{eq: hamiltonian without MF coupling}
\end{align}
where $J >0$ is the ferromagnetic coupling constant $\bm{B}(t)$ is an external magnetic field ($g$ is  Lande's $g$ factor and $\mu_B$ is the Bohr magneton). 
The sum $\sum_{\langle \bmr, \bmr\pri \rangle}$ is taken over all the pairs of the nearest-neighbor sites.
We measure the external magnetic field in units of $g\mu_B$ and thereby set $g \mu_B = 1$. 
We apply a circularly polarized driving magnetic field with strength $B_d$ in the $m_x$-$m_y$ plane 
[see Fig.~\ref{fig: sLLG_setup} (b)]. 
The total field $\bm{B}(t)$ is given by 
\begin{align}
	\bm{B}(t) &= \br{
	B_s + B_d \cos(\omega t), 
	- B_d \sin(\omega t), 
	0
	}^{\mr{tr}},  \label{eq: sLLG, external, field} 
\end{align}
which is decomposed into the Fourier harmonics as follows:
\begin{align}
	\bm{B}(t) &= \bm{B}_0 + \bm{B}_1 \Exp{- i\omega t} + \bm{B}_{-1} \Exp{i\omega t}
	, \nonumber \\
	\bm{B}_0 &= (B_s, 0, 0)^{\mr{tr}}
	, \nonumber \\
	\bm{B}_{\pm 1} &= 
	{B_d \over 2} (1, \mp i, 0)^{\mr{tr}}
	. 
\end{align}
For real magnetic materials, the typical value of the exchange interaction $J$ is the order of $1$ - $10$meV. 
Therefore, the frequency $\omega$ of applied electromagnetic waves should be the same as the energy scale of $J$, corresponding to the range from gigahertz to terahertz.
The model \eqref{eq: hamiltonian without MF coupling} would be relevant for the laser-driven spin dynamics in ordered magnets. 
Note that a quantum analog of the model \eqref{eq: hamiltonian without MF coupling} with an anisotropic term has been studied in Refs.~\cite{TakayoshiS14a, TakayoshiS14b}.

\subsection{FM expansion and the effective sLLG}
The Fourier harmonics $\bmf_{\pm1}$ and the resultant first-order FM drift force $\bmf_{F, \bmr}^{(1)}$ are obtained from Eqs.~\eqref{eq: FME, 1st} and \eqref{eq: sllg, fg} as follows (see App.~\ref{app: der, sLLG} for the derivation): 
\begin{align}
	\bmf_{\pm 1, \bmr} &=
	 - {\bmm_{\bmr} \over 1 + \alpha^2} 
	 \times 
	\br{ \bm{B}_{\pm 1} + {\alpha \over m_s} \bmm_{\bmr} \times \bm{B}_{\pm 1}}
	, \nonumber \\
	\bmf_{F, \bmr}^{(1)}
	 &= {i \over \omega} \rbr{
	\bmf_{1, \bmr} \cdot {\delta \bmf_{-1, \bmr} \over \delta \bmm_{\bmr}}
	 - \bmf_{-1, \bmr} \cdot {\delta \bmf_{1, \bmr} \over \delta \bmm_{\bmr}}
	} \nonumber \\
	 &= - {\bmm_{\bmr} \over 1 + \alpha^2} 
	 \times 
	\br{ \bm{H}_{F, \bmr}^{(1)} + {\alpha \over m_s} \bmm_{\bmr} \times \bm{H}_{F, \bmr}^{(1)}}
	, \nonumber \\
	\bm{H}_{F, \bmr}^{(1)} &= 
	{i \bm{B}_{-1} \times \bm{B}_{+1} \over (1 + \alpha^2) \omega} - \alpha {i \bm{B}_{-1} \times \bm{B}_{+1} \over (1 + \alpha^2) \omega} \times \bmm_{\bmr}
	. \label{eq: sLLG, FM, 1stfield}
\end{align}
While the first term in $\bm{H}_{F, \bmr}^{(1)}$ represents the effective magnetic field with 
\begin{align}
	\bm{b}^{(1)} := {i \bm{B}_{-1} \times \bm{B}_{+1} \over (1 + \alpha^2) \omega} 
	 = {(B_d)^2 \over 2\omega(1+ \alpha^2)} \hat{z}
	, \label{eq: induced B}
\end{align}
the second term describes the so-called spin-transfer torque \cite{SlonczewskiJ96, BergerL96, TsoiM98, MyersE99}. 
The effective field parallel to $i (\bm{B}_{-1} \times \bm{B}_{+1})$ is the dominant term of the FM expansion, 
and thereby we can predict that the circularly polarized laser changes the value of the magnetization parallel to $i (\bm{B}_{-1} \times \bm{B}_{+1})$. 
The emergence of this effective magnetic field can be qualitatively understood from an analog with a quantum system: 
in the presence of the external drive $\hat{H}(t) = -\bm{B}(t) \cdot \hat{\bm{S}}$, with $\hat{\bm{S}}$ being the total spin operator in the quantum system, the first-order FM Hamiltonian 
\begin{align}
	\hat{H}_F^{(1)} 
	:= { [\hat{H}_{-1}, \hat{H}_1] \over \omega }
	= {i \over \omega} 
	\br{ \bm{B}_{-1} \times \bm{B}_{+1} }
	\cdot \hat{\bm{S}}
\end{align}
represents the effective magnetic field $i (\bm{B}_{-1} \times \bm{B}_{+1} ) / \omega$ \cite{TakayoshiS14a, TakayoshiS14b}.
However, we find from Eqs.~\eqref{eq: sLLG, FM, 1stfield} that its magnitude decreases by a factor of $(1 + \alpha^2)^{-1}$. 
Moreover, there appears the spin-transfer torque as a consequence of the coupling with the environment, which is absent in an isolated spin system. 
The drift field $\bmf_{\mr{mic}, \bmr}^{(1)}$ corresponding to the kick operator $\mG_F^{(1)}(t)$ in Eq.~\eqref{eq: kick, 1st} is given by 
\begin{align}
	& \bmf_{\mr{mic}, \bmr}^{(1)}(\bmm_{\bmr}, t)
	 := 
	\nonumber \\
	& - {\bmm_{\bmr} \over 1 + \alpha^2} 
	 \times 
	\br{ \bm{H}_{\mr{mic}}^{(1)}(t) + {\alpha \over m_s} \bmm_{\bmr} \times \bm{H}_{\mr{mic}}^{(1)}(t)}
	, 
\end{align}
where $\bm{H}_{\mr{mic}}^{(1)}(t) = i \br{\bm{B}_{-1} \Exp{i \omega t} - \bm{B}_{+1} \Exp{- i \omega t}}$ describes an oscillating magnetic field.

The second-order-expansion terms are much more complicated due to the renormalization of the diffusion matrix $G$. 
The effective sLLG equation is obtained to be
\begin{align}
	\dot{\bmm}_{\bmr} & = 
	 - {\bmm_{\bmr} \over 1 + \alpha^2 } 
	 \times 
	\rbr{ \bm{H}_{F, \bmr} + \sqrt{1+\chi_{\bmr}}\bmh_{\bmr} 
	\right.
	\nonumber \\
	&\quad 
	\left.
	 + {\alpha \over m_s} \bmm_{\bmr} \times 
	 \br{ \bm{H}_{F, \bmr} + \sqrt{1+\chi_{\bmr}} \bmh_{\bmr} }}
	, \label{eq: eff, sLLG, 2nd}
\end{align}
where the effective magnetic field $\bm{H}_{F, \bmr}$ and the correction $\chi_{\bmr}(\bmm_{\bmr})$ to the diffusion term are given as follows (see App.~\ref{app: der, sLLG} for the derivation):  
\begin{align}
	\bm{H}_{F, \bmr} &:= \sum_{\bmr\pri: \mr{n.n.}} 
	\br{ J \bmm_{\bmr\pri} + \bm{\delta}\bm{J}_{\bmr, \bmr\pri} }
	  + \bm{B}_F + \bm{V}_F \times \bmm_{\bmr}
	, \nonumber \\
	\chi_{\bmr}(\bmm_{\bmr}) &:=
	- \br{{\alpha B_d \over m_s \omega (1 + \alpha^2)}}^2 
	{ 3(m_s)^2 - (m_{\bmr,z})^2 \over 2}
	. \label{eq: effective field FM2}
\end{align}
Here, the sum $\sum_{\bmr\pri: \mr{n.n.}}$ is taken over the nearest-neighbor sites of $\bmr$. 
The total effective external magnetic field $\bm{B}_F$, the spin-transfer torque $\bm{V}_F$, and the effective interaction $\bm{\delta}\bm{J}_{\bmr, \bmr\pri}$ are given by 
\begin{align}
	&\bm{B}_F :=
	\bm{B}_0 + 
	\bm{b}^{(1)} + (1 - \alpha^2) \bm{b}^{(2)} 
	- {\alpha m_s D \over 2(1+ \alpha^2) } 
	{\delta \chi_{\bmr} \over \delta \bmm_{\bmr}}
	, \nonumber \\ 
	&\bm{V}_F :=
	 - {\alpha \over m_s} \bm{b}^{(1)} 
	 - {2 \alpha \over m_s} \bm{b}^{(2)}
	  + { D  \over 2(1+ \alpha^2 )} 
	  {\delta \chi_{\bmr} \over \delta \bmm_{\bmr}}
	  ,
	\nonumber \\
	& \bm{\delta}\bm{J}_{\bmr, \bmr\pri} := 
	J \br{{\alpha B_d \over m_s \omega (1 + \alpha^2)}}^2
	\nonumber \\
	& \times
	\sum_{\bmr\pri: n,n} m_{\bmr\pri, z}
	\thvec{m_{\bmr\pri, x} \delta m_{\bmr, \bmr\pri,z}}
	{m_{\bmr\pri, y} \delta m_{\bmr, \bmr\pri,z}}
	{- m_{\bmr\pri, z} \delta m_{\bmr, \bmr\pri,x} - 
	m_{\bmr\pri, y} \delta m_{\bmr, \bmr\pri,y}}
	, 	  
\end{align}
where $\delta \bmm_{\bmr, \bmr\pri} := \bmm_{\bmr} - \bmm_{\bmr\pri}$ and $\bm{b}^{(2)}$ is defined by 
\begin{align}
	\bm{b}^{(2)} & := 
	- \br{{ B_d \over 2 \omega (1 + \alpha^2)}}^2 B_s \hat{x}
	. 
\end{align}


\subsection{Short-time dynamics}
In Fig.~\ref{fig: sLLG_setup} (c), 
we calculate the time evolution of the spatially averaged magnetization 
$S_z := (1 / N) \sum_{\bmr} m_{\bmr, z}$, with $N = 100 \times 100$ being the number of spins, 
using three different equations: 
(i) the original sLLG equation \eqref{eq: sLLG} with driving field (red), 
(ii) the effective one \eqref{eq: eff, sLLG, 2nd} without the kick operator $\mG_F$ (blue), and 
(iii) the effective one with the kick operator $\mG_F$ (green). 
We use the Heun method for numerical integration of the sLLG equation with the linearization technique \cite{SekiS16}.  

The initial state is taken as the fully polarized state: $\bmm_{\bmr} = (1,0,0)^{\mr{tr}}$ for all the cases. 
The parameters are chosen as $J = 1, B_d = 1.4 J, \omega = 5 J, B_s = 1.4 J, \alpha  = 0.1$, and $T_{\mr{te}} = 0$. 
After a long time $t \gg (\alpha J)^{-1}$, the system approaches the NESS, where the magnetization oscillate with period $T$.
Due to the effective magnetic field $\bm{b}_F^{(1)} (\parallel \hat{z})$, the average magnetization $S_z$ takes a positive value. 
As we can see from Fig.~\ref{fig: sLLG_setup} (c), the effective sLLG equation with the kick operator (blue) well approximates the exact time evolution (red). 
Although the effective sLLG equation without the kick operator (green) fails to capture the oscillating behavior, it correctly reproduces the long-time average $\bar{S}_z$ for the steady state.

\begin{figure}[t!]
\centering
   \includegraphics[width=\columnwidth, clip]{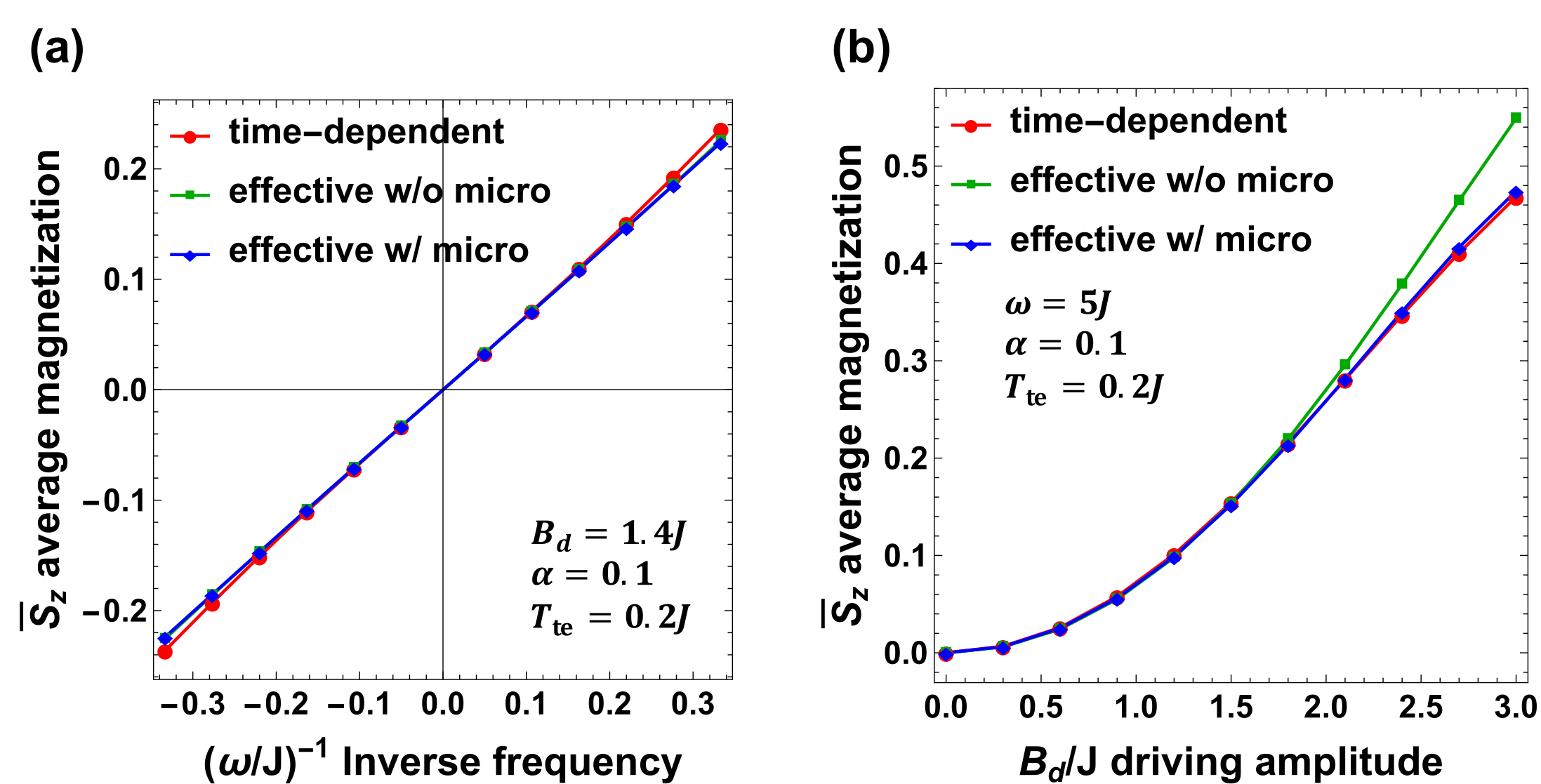}\\
   \includegraphics[width=\columnwidth, clip]{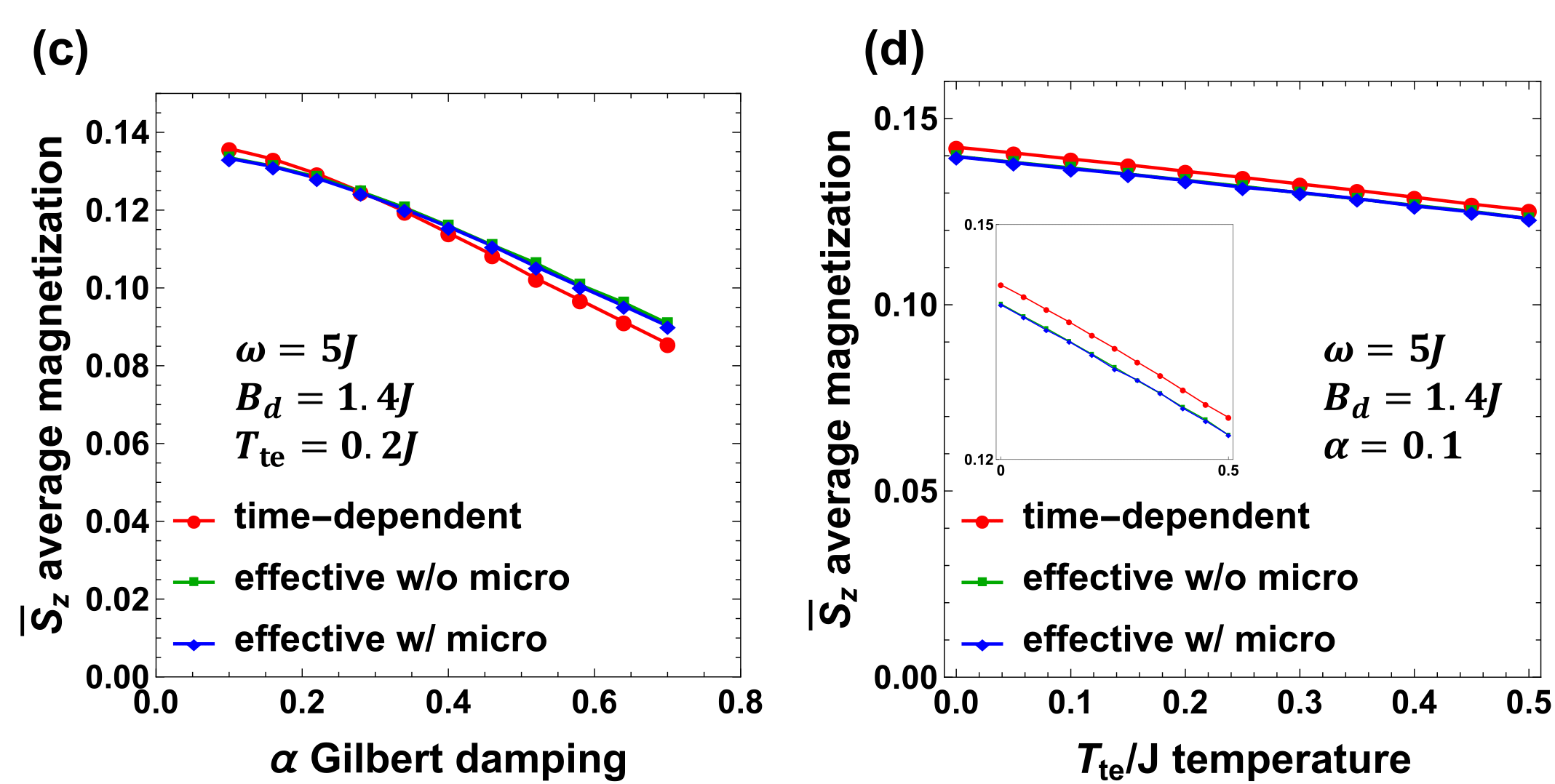}
\caption{Dependence of the long-time average of the magnetization $\bar{S}_z$ on (a) the driving frequency $\omega$, (b) the driving strength $B_d$, (c) the Gilbert damping $\alpha$, and (d) temperature $T_{\mr{te}}$.  
	The red, green, and blue points are obtained from the time-dependent sLLG equation \eqref{eq: sLLG}, the effective one \eqref{eq: eff, sLLG, 2nd} without the kick operator, and that with the kick operator, respectively. 
	The parameters are fixed as $J = 1, \omega = 5 J, B_d = 1.4 J, B_s = 1.4 J, \alpha  = 0.1$, and 
$T_{\mr{te}} = 0.2 J$, except for the parameter that is varied in each panel. The magnetization is normalized as $m_s=1$. 
	The inset in the panel (d) shows an enlarged image between $0.12 \le \bar{S}_z \le 0.15$. 
    }
       \label{fig: sLLG_ness}
\end{figure}

\subsection{NESS}
In Fig.~\ref{fig: sLLG_ness}, 
we show a comprehensive analysis of the dependence of the long-time average of the magnetization $\bar{S}_z$ on (a) the driving frequency $\omega$, (b) the driving amplitude $B_d$, (c) the Gilbert damping $\alpha$, and (d) temperature $T_{\mr{te}}$.  
The curves with the three colors, red, green, and blue, are obtained from the three equations (i), (ii), and (iii), respectively. 
Except for the parameter that is changed in each panel, the parameters are fixed as $J = 1, \omega = 7 J, B_d = J, B_s = 1.4 J, \alpha  = 0.1$, and $T_{\mr{te}} = 0.2 J$. 
As shown in Fig.~\ref{fig: sLLG_ness} (a), 
the average magnetization $\bar{S}_z (\propto \omega^{-1})$ is induced by the effective magnetic field $\bm{b}^{(1)} (\propto \omega^{-1})$, and the time-dependent sLLG equation \eqref{eq: sLLG} and the effective one \eqref{eq: eff, sLLG, 2nd} show excellent agreement in the high-frequency regime $\omega / J >5$. 
In Fig.~\ref{fig: sLLG_ness} (b), the driving amplitude is varied from weakly driven ($B_d \ll J$) to strongly driven ($B_d \simeq J$) regimes, where the effective sLLG equation with kick operator shows better agreement for a strong drive. 
The effect of the Gilbert damping is analyzed in Fig.~\ref{fig: sLLG_ness} (c), where the calculation is performed from weakly dissipative ($\alpha \ll 1$) to strongly dissipative ($\alpha \simeq 1$) regimes. 
In Fig.~\ref{fig: sLLG_ness} (d), we vary the temperature to simulate the sLLG equation with ($T_{\mr{te}} = 0$) and without ($T_{\mr{te}} > 0$) the random field $\bm{h}_{\bmr}$. 
Although a slight deviation is visible between the time-dependent and effective sLLG equation (see the inset), the latter correctly well reproduces the temperature dependence, i.e., the slope of the curve. 
From the above results, we can conclude the effective sLLG equation obtained from the FM expansion correctly reproduces the exact time evolution in a wide range of parameters from weak to strong dissipation and with and without a random field. 
These numerical results support the analytical arguments on the validity of the FM expansion in Sec.~\ref{sec: conv}.

\begin{figure}[t!]
\centering
   \includegraphics[width=\columnwidth, clip]{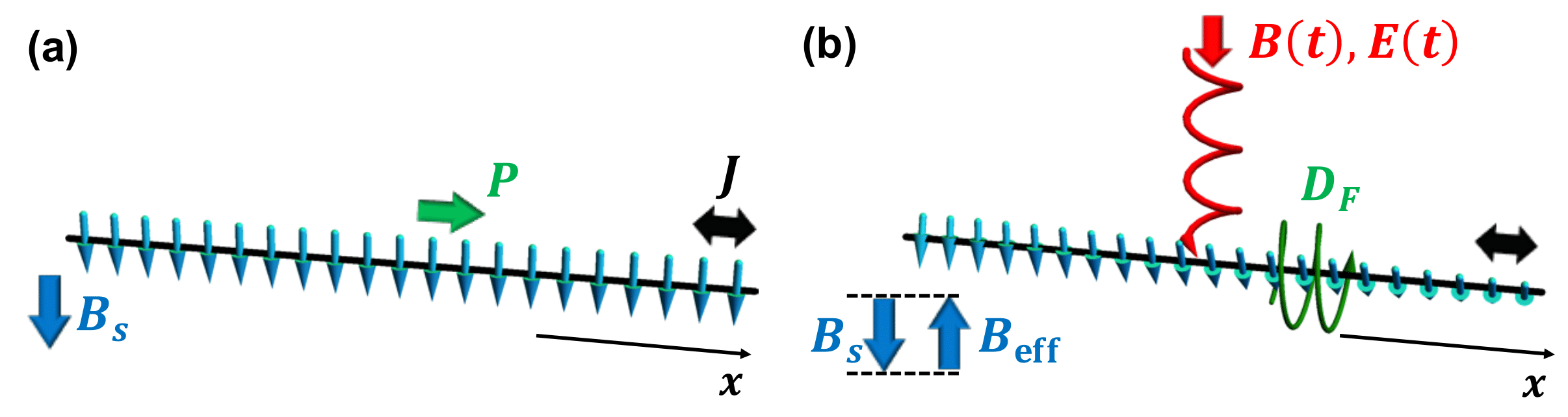}
\caption{
(a) Schematic illustration of a multiferroic spin chain. 
The nearest-neighbor spins are coupled with a ferromagnetic coupling with exchange coupling $J$ and the ME coupling with polarization $\bm{P}$ given by Eq.~\eqref{invDM}, and a static field $\bm{B}_s$ is applied in the $m_z$ direction. 
(b) By the irradiation of the laser with electric fields $\bm{E}(t)$ and magnetic field $\bm{B}(t)$, the effective DM interaction $\bm{D}_F$ and the effective magnetic field $\bm{D}_F$ are induced, 
leading to a spiral spin texture and an emergence of the vector chirality $\mathcal{V}_x^{\mr{tot}}$ 
along the $x$ direction. The static field is tuned to cancel the effective field $\bm{B}_F$.
    }
       \label{fig: sLLG_MF_setup}
\end{figure}  

\section{\label{sec: spintronics}Application to spintronics}
Laser control of magnetic materials attracts considerable interest in recent years since it could offer an ultrafast and non-contact manipulation of magnets \cite{KirilyukA10}. 
As demonstrated in Refs.~\cite{TakayoshiS14a, TakayoshiS14b, SatoM16, ClaassenM17, KitamuraS17, MentinkJ15}, lasers can serve as a versatile tool for a dynamical control of their magnetic moment and even their intrinsic interactions, e.g., exchange interaction and Dzyaloshinskii-Moriya (DM) interaction. 
Multiferroics are materials that exhibit both ferromagnetism and ferroelectricity \cite{TokuraY14, WangK09}.
Due to the coupling between the spin degrees of freedom and the electric polarization, they have potential applications to future spintronics \cite{MaekawaS17}. 

In this section, we consider a multiferroic spin system described by the sLLG equation. 
Our setup is a classical analog of Ref.~\cite{SatoM16}, where a periodically driven multiferroic \textit{quantum} spin chain was studied. 
While a vector spin chirality and a spin current have been shown to emerge in the previous study, 
they vanish after a long time due to heating. 
On the other hand, in our setup, the system reaches the NESS with a finite vector chirality 
due to the balance between the driving and the Gilbert damping. 

\subsection{Synthetic DM interaction in a multiferroic spin chain}
The Hamiltonian $\mathcal{H}_{\mr{MF}}(t)$ is given from $\mathcal{H}(t)$ in Eq.~\eqref{eq: hamiltonian without MF coupling} by adding the magnetoelectric (ME) coupling: 
\begin{align}
	\mathcal{H}_{\mr{MF}}(t) = 
	\mathcal{H}(t) - \bm{P} \cdot \bm{E}(t)
	, 
\label{eq:MultiferroModel}
\end{align}
where the second term represents the ME coupling of the total polarization $\bm{P}$ with an external electric field $\bm{E}(t)$. 
We assume the antisymmetric magnetostriction mechanism (also known as the inverse DM effect mechanism  \cite{KatsuraH05, KatsuraH07, KatsuraH05, SergienkoI06, MostovoyM06}) of mutliferroics, where $\bm{P}$ is coupled with a local magnetization $\bmm_{\bmr}$ through the vector spin chirality $\bmm_{\bmr} \times \bmm_{\bmr\pri}$:
\begin{align}
	\bm{P} 
	& = \sum_{\langle \bmr, \bmr\pri \rangle} \bm{P}_{\bmr, \bmr\pri} 
	 = g_{me}
	 \sum_{\langle \bmr, \bmr\pri \rangle} 
	 \bm{e}_{\bmr, \bmr\pri} \times 
	\br{\bmm_{\bmr} \times \bmm_{\bmr\pri} }
	. \label{invDM}
\end{align}
Here, $\bm{e}_{\bmr, \bmr\pri} := (\bmr\pri - \bmr)/|\bmr\pri - \bmr|$ is the unit vector connecting the nearest-neighbor sites $\bmr$ and $\bmr\pri$, and $g_{me}$ denotes the magnitude of the ME coupling. 
This ME coupling is known to be responsible for electric polarization in a wide class of spiral ordered (i.e., chirality ordered) multiferroic magnets. 
Combining Eqs.~\eqref{eq: hamiltonian without MF coupling} and \eqref{invDM}, we obtain the explicit forms of $\mathcal{H}_{\mr{MF}}(t)$ and the effective field $\bm{H}_{\bmr}(t)$:
\begin{align}
	\mathcal{H}_{\mr{MF}}(t) 
	&= - 
	\sum_{\langle \bmr, \bmr\pri \rangle} 
	\rbr{
	J \bmm_{\bmr} \cdot \bmm_{\bmr\pri}
	 + \bm{D}_{\bmr, \bmr\pri}(t) \cdot 
	\br{\bmm_{\bmr} \times \bmm_{\bmr\pri}}
	}
	\nonumber \\
	 &\quad - \sum_{\bmr} 
	\rbr{\bm{B}_s + \bm{B}(t)} \cdot \bmm_{\bmr} 
	, \nonumber \\
	\bm{H}_{\bmr}(t) &= \sum_{\bmr\pri: n.n.} 
	\br{
	J \bmm_{\bmr\pri}
	 + \bm{D}_{\bmr, \bmr\pri}(t) \times \bmm_{\bmr\pri}
	} + \bm{B}_s + \bm{B}(t)
	, \label{eq: mf, eff_field, ori}
\end{align}
where $\bm{D}_{\bmr, \bmr\pri}(t) := g_{me} \bm{E}(t) \times \bm{e}_{\bmr, \bmr\pri}$ is the DM coupling induced by the electric field. 
To demonstrate the emergence of a spin texture by a laser, we consider a spin chain aligned along the $x$ direction irradiated by the laser field traveling along the $(-z)$ direction (see Fig.~\ref{fig: sLLG_MF_setup} (a)). 
Although a realistic multiferroic system has a strong three-dimensional nature \cite{MiyadaiT83, ShinozakiM16}, we here consider a one-dimensional chain for simplicity.
The electric field $\bm{E}(t)$ and the magnetic field $\bm{B}(t)$ of an applied circularly-polarized electromagnetic wave are given by 
\begin{align}
	\bm{E}(t) &= E_0 \br{
	\sin(\omega t), \cos(\omega t), 0 
	}^{\mr{tr}}
	, \nonumber \\
	\bm{B}(t) &= { - \hat{z} \over c} \times \bm{E}
	 = 
	{E_0 \over c} \br{
	\cos(\omega t), -\sin(\omega t), 0 
	}^{\mr{tr}}
	. \label{eq: mf, laser field} 
\end{align}

From the first-order FM expansion, 
the effective static field $\bm{H}_{\bmr, F}$ at site $\bmr$ is obtained from Eq.~\eqref{eq: FME, 1st} to be
\begin{align}
	\bm{H}_{\bmr, F} 
	:= &
	\sum_{\bmr\pri: n.n.} 
	\rbr{
	J \bmm_{\bmr\pri}
	 + \bm{D}_{F, \bmr, \bmr\pri} \times \bmm_{\bmr\pri}
	} + \bm{B}_F + \bm{B}_s
	\nonumber \\
	 & - \alpha \bm{B}_F \times \bmm_{\bmr}
	 \nonumber \\
	& - \sum_{\bmr\pri: n.n.} 
	{\alpha \epsilon_B\epsilon_E \over 2 m_s (1+  \alpha^2) \omega}
	  \thvec{ m_s^2 + m_{\bmr\pri, y}\delta m_{\bmr, y}}{- m_{\bmr\pri, y}\delta m_{\bmr, x}}{0}
	 , \nonumber \\
	 \bm{D}_{F, \bmr, \bmr\pri} &=
	{ \epsilon_E \epsilon_B \over  2 ( 1 + \alpha^2) \omega } \bm{e}_{\bmr, \bmr\pri}
	=: D_F \bm{e}_{\bmr, \bmr\pri}
	, \nonumber \\
	\bm{B}_F &= { \epsilon_{B}^2 \over 2 ( 1 + \alpha^2) \omega} \hat{z}
	, \label{eq: mf, eff_field, FM}
\end{align}
where $\epsilon_E := g_{me} E_0$ and $\epsilon_{B} := (g \mu_B E_0)/c$ are the normalized electric and magnetic energies, respectively (see App.~\ref{app: der, sLLG} for the derivation). 
Equation \eqref{eq: mf, eff_field, FM} shows that a synthetic DM field $\bm{D}_{F, \bmr, \bmr\pri}$ emerges from the combination of the ME and Zeeman couplings. 
The strongest magnetic field $\epsilon_B$ of terahertz  lasers attains $1$ - $10$ T \cite{HiroiH11, PashkinA13} and the magnitude of $g_{me}$ can be large in a gigahertz to terahertz region \cite{TakahashiY11, HuvonenD09, KatsuraH07, FurukawaS10}. 
For standard magnets with $J = 0.1$ - $10$ meV, 
both $\epsilon_E / J$ and $\epsilon_B / J$ can achieve values of $0.1$ - $1$. 
Second and third terms in $\bm{H}_{\bmr, F}$ are the laser-driven DM interaction generated via the single-photon absorption and emission, and the synthetic magnetic field appeared in Sec.~\ref{sec: sLLG}, respectively.

We can ignore the second and third lines of $\bm{H}_{\bmr, F}$ in Eq.~\eqref{eq: mf, eff_field, FM} provided that the dissipation is weak enough ($\alpha \ll 1$), where the resulting sLLG equation is equivalent to the sLLG equation with the static Hamiltonian 
\begin{align}
	\mathcal{H}_{F} = &
	- \sum_{j=1}^L \rbr{
	J \bmm_j \cdot \bmm_{j+1}
	 + D_F \mathcal{V}_{j, x} +  (\bm{B}_F + \bm{B}_s) \cdot \bmm_j
	}
	.
	\label{eq: eff, VC}
\end{align}
Here $\mathcal{V}_{j, x} := \hat{x} \cdot (\bmm_j \times \bmm_{j+1})$ is the vector chirality along the $x$ axis. 
It is clear from Eq.~\eqref{eq: eff, VC} that the system exhibits a spiral spin texture due to the synthetic DM interaction, leading to the emergence of the vector chirality (see Fig.~\ref{fig: sLLG_MF_setup} (b)). 
To maximize the total vector chirality $\mathcal{V}^{\mr{tot}}_x := \sum_j \mathcal{V}_{j, x}$, we introduce a static field $\bm{B}_s$ along the $z$ axis to cancel out $\bm{B}_F$, i.e., $\bm{B}_s + \bm{B}_F = 0$. 
With purely ferromagnetic and DM interactions in Eq.~\eqref{eq: eff, VC}, a spin spiral state emerges \cite{KishineJ05, TogawaY13}, whose vector chirality per site is given as follows: 
\begin{align}
	{\mathcal{V}^{\mr{tot}}_{x} \over L} &:=
	{1 \over L} \sum_j \mathcal{V}_{j, x} 
	= \tan^{-1}\br{{D_F \over J}}
	\nonumber \\
	&= \tan^{-1}\rbr{{\epsilon_E \epsilon_B \over  2 ( 1 + \alpha^2) \omega J}}
	. \label{eq: mf, vc, theory}
\end{align}

\begin{figure}[t]
\centering
   \includegraphics[width=\columnwidth, clip]{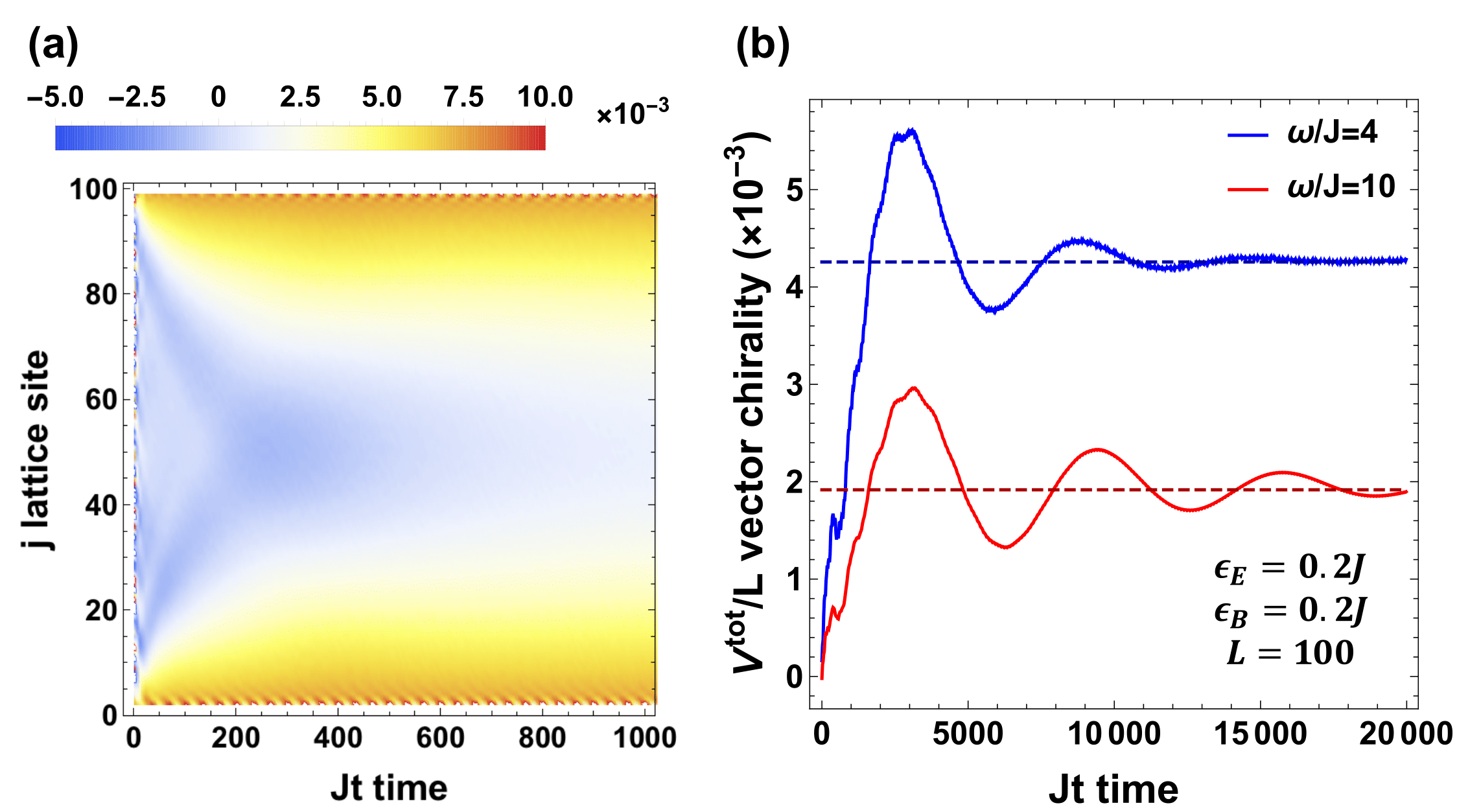}\\
   \includegraphics[width=\columnwidth, clip]{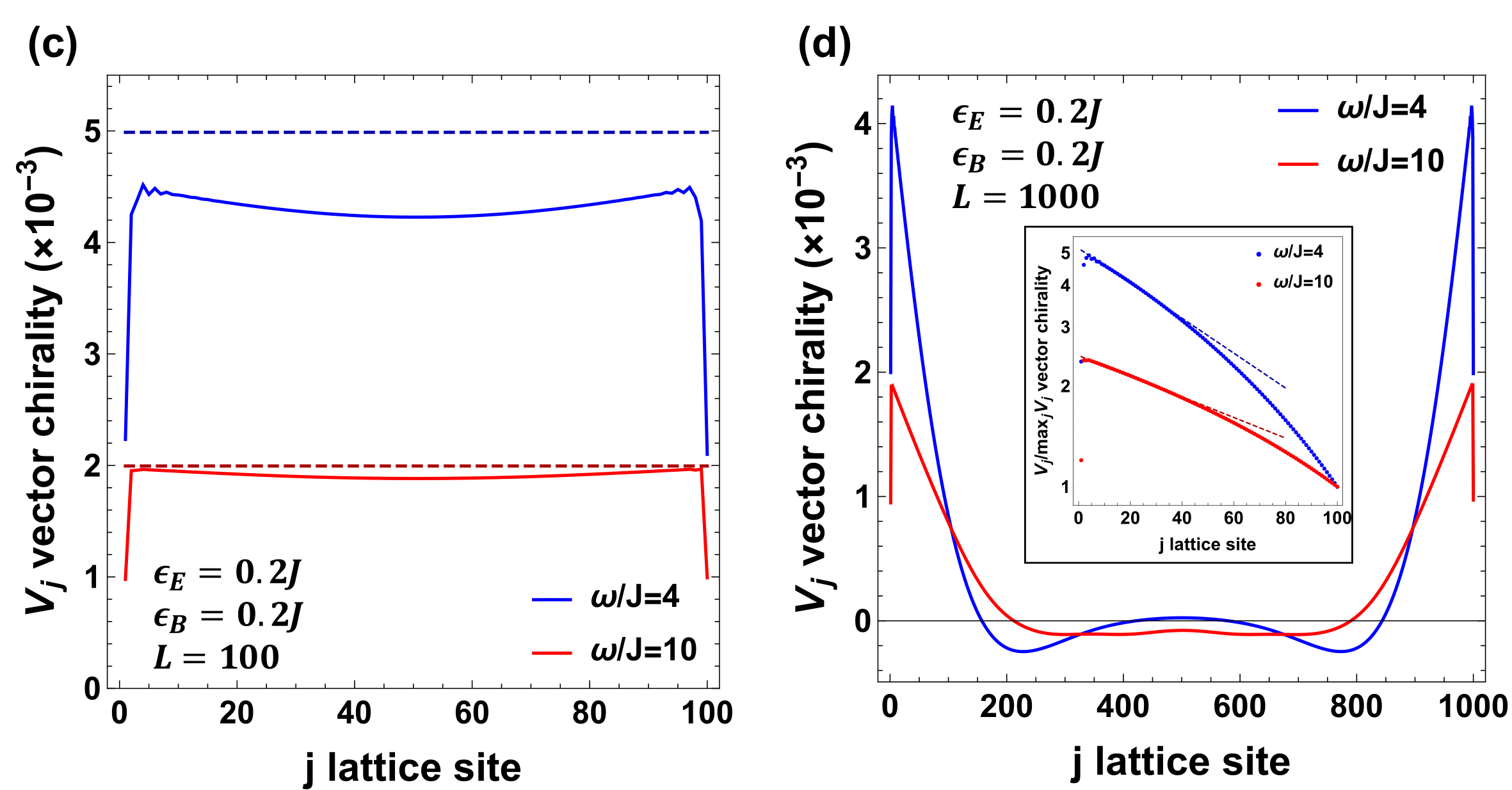}
\caption{
	(a): Spatiotemporal dynamics of the vector chirality $[\hat{x} \cdot (\bmm_j \times (\bmm_{j+1} -\bmm_{j-1}))] / 2$ with driving frequency $\omega/J = 4$. 
	As time passes, the vector chirality enters the system from the edges $j = 0, L$, where it spreads uniformly over the system after a sufficiently long time. 
	(b): Time evolution of the spatially averaged vector chirality $\mathcal{V}^{\mr{tot}}_x / L$ with different driving frequencies $\omega/J = 4$ (blue) and $\omega/J = 10$ (red). The dashed lines shows the values at the NESSs.
	(c), (d): Spatial profile of $\mathcal{V}_{j, x}$ at NESS for two system sizes [$L = 100$ for (c) and $L = 1000$ for (d)] in the NESSs. The inset in Fig. (d) shows the logarithmic plot of $\mathcal{V}_{j, x} / (\mr{max}_j \mathcal{V}_{j, x})$ for the first 100 sites $0 \le j \le 100$. 
	The red and blue points and curves corresponds to the smaller ($\omega = 4 J$) and larger ($\omega = 10 J$) driving frequency. 
    }
       \label{fig: vc, 1}
\end{figure}

\subsection{Emergent vector chirality by laser irradiation}

To demonstrate the emergent vector chirality $\mathcal{V}^{\mr{tot}}_x$ predicted 
from the effective-theory analysis \eqref{eq: mf, vc, theory}, 
we perform a numerical simulation of the time-dependent sLLG equation \eqref{eq: sLLG, original} with time-dependent effective magnetic field $\bm{H}_{\bmr}(t)$ \eqref{eq: mf, eff_field, ori}. 
We fix the Zeeman coupling $\epsilon_{B}$, the Gilbert damping, and the temperature 
$T_{\mr{te}}$ as $\epsilon_{B}/J = 0.2$, $\alpha = 0.05$, and $T_{\mr{te}}  =0$, respectively. 
The initial state is set to be the polarized state $\bmm_{\bmr} = - \hat{z}$, and the laser is turned on at $t = 0$. 
Since $\mathcal{V}_{j, x}$ emerges from the edges as we will see below,  we solve the sLLG equation with the open boundary condition, i.e., $\bmm_0 = \bmm_{L+1} = \bm{0}$, 
In Fig.~\ref{fig: vc, 1} (a) shows the spatiotemporal dynamics of the vector chirality 
$[\hat{x} \cdot (\bmm_j \times (\bmm_{j+1} -\bmm_{j-1}))] / 2$, while the time evolution of the spatially averaged vector chirality $\mathcal{V}^{\mr{tot}}_x / L$ is plotted in Fig.~\ref{fig: vc, 1} (b). 
As we can see from Fig.~\ref{fig: vc, 1} (a), the vector chirality enters the system from the edges ($j = 0, L$) in the initial relaxation ($t \lesssim 200 J^{-1}$) and spreads uniformly over the system after a sufficiently long time ($t \sim 10^4 J^{-1}$). 
Due to the balance between the drive and the damping, the system reaches the NESS with constant 
$\mathcal{V}^{\mr{tot}}_x / L$ [dashed lines in Fig.~\ref{fig: vc, 1} (b)]. 
In Figs.~\ref{fig: vc, 1} (c) and (d), we plot the spatial profiles of $\mathcal{V}_{j, x}$ 
at the NESSs for the chain lengths $L = 100$ and $L = 1000$, respectively. 
The vector chirality in the NESSs is localized at the edges as shown in Fig.~\ref{fig: vc, 1} (d). 
However, the vector chirality uniformly spreads over the system for a smaller system ($L = 100$) as shown in Fig.~\ref{fig: vc, 1} (c) due to a rather long  localization length (e.g., $\sim 100$ sites for $\omega/J = 4$). 
Note that the localization length becomes even larger for larger $\omega$ as shown in the inset in Fig.~\ref{fig: vc, 1} (d). 
This implies that one can optically induce a tunable vector chirality for nanomagnets and disordered spin systems where impurities effectively play the role of boundaries.

\begin{figure}[t]
\centering
\includegraphics[width=\columnwidth, clip]{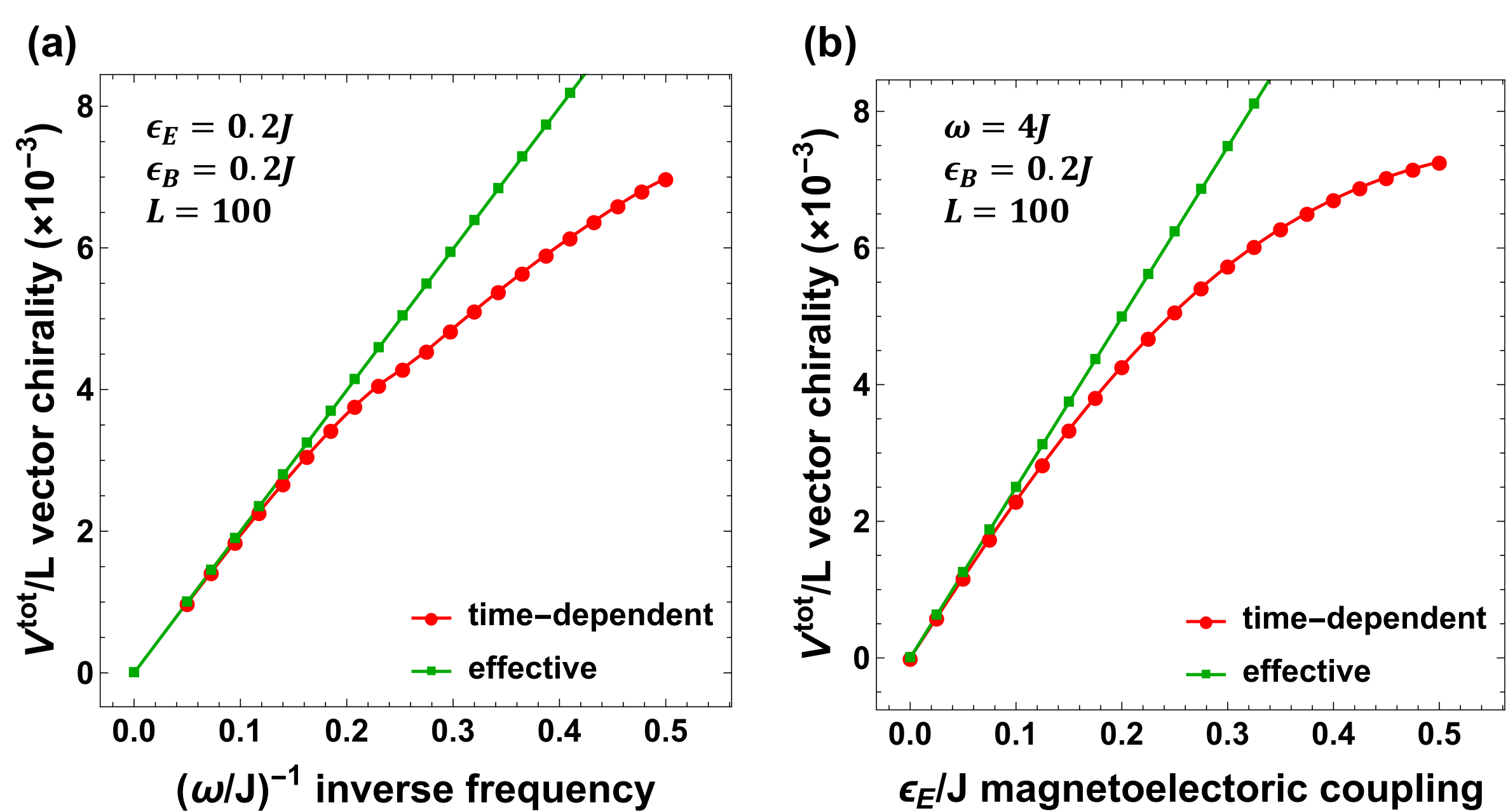}
\caption{
	(a) Dependence of the mean vector chirality $ \mathcal{V}^{\mr{tot}}_x / L$ 
on the driving frequency $\omega$ in the NESS with a fixed ME coupling $\epsilon_E = 0.2 J$. 
	(b) Dependence of the mean vector chirality $ \mathcal{V}^{\mr{tot}}_x / L$ 
on the ME coupling $\epsilon_E$ in the NESS with a fixed driving frequency $\omega = 4 J$. 
	The red curves are obtained from the solution of the sLLG equation with a time-dependent effective field \eqref{eq: mf, eff_field, ori} while green ones are drawn from Eq.~\eqref{eq: mf, vc, theory} that is derived from the FM expansion. 
    }
       \label{fig: vc, 2}
\end{figure}  

Finally, we quantitatively check the validity of our effective theory by calculating the dependence of the mean vector chirality $\mathcal{V}^{\mr{tot}}_x / L$ at the NESS on the frequency $\omega$ and the ME coupling $\epsilon_E$ to compare with the effective-theory result \eqref{eq: mf, vc, theory}. 
As shown in Fig.~\ref{fig: vc, 2}, we have excellent agreement between the exact NESS and the effective-theory analysis \eqref{eq: mf, vc, theory} in the high-frequency or weak ME-coupling regions, 
where the synthetic DM interaction $D_F$ is small and hence the FM expansion is expected to be good. 
This result is consistent with the analytical one on the validity of the FM expansion in Sec.~\ref{sec: conv}.

\section{\label{sec: summary}Conclusion and outlook}
In this paper, we develop the FM expansion of periodically driven classical EOMs. 
Our formalism is applicable not only to classical systems but also to quantum ones and to both isolated and open ones at zero and finite temperatures, as long as they are described by nonlinear (stochastic) differential equations. 
The key idea is using the master equation corresponding to the EOM to which we apply the Floquet theorem and perform the FM expansion (see Fig.~\ref{fig: strategy}). 
The FM expansion of the EOM is obtained from that of the time-periodic master equation. 
By analytical evaluation of the higher-order terms of the FM expansion, we find that it is, at least asymptotically, convergent and well reproduces a NESS for a systems. 
Our method is demonstrated in a single particle system and a many-body system by examples of a Kapitza pendulum with friction (Sec.~\ref{sec: Kapitza}) and laser-driven magnets described by the sLLG equation (Sec.~\ref{sec: sLLG}), respectively. 
In both cases, the effective EOM obtained from the FM expansion is found to well approximate the time-dependent one not only in a short time during an initial relaxation but also for a long time up to their NESSs. 
This result is in stark contrast to isolated systems where the truncated FM expansion fails to capture the heating to an infinite-temperature state after the Floquet prethermalization. 
Finally, in Sec.~\ref{sec: spintronics}, we present an application to spintronics, demonstrating an optical generation of a spin vector chirality in a multiferroic spin chain by a circularly polarized laser.

This work opens many avenues for future exploration. 
First, it will be interesting to study an application to ultrafast spintronics \cite{KirilyukA10}. 
While Floquet engineering of magnets is mostly discussed in isolated systems \cite{TakayoshiS14a, TakayoshiS14b, SatoM16, ClaassenM17, KitamuraS17, MentinkJ15}, the coupling with an environment is unavoidable in any solid-state system. 
Also, numerical simulations of interacting spin systems are restricted to a small system size due to the exponentially increasing dimension of the Hilbert spaces, where a finite-size effect is inevitable. 
In contrast, by means of our approach, one can simulate driven classical spin systems with considerably larger system sizes than quantum ones, yet to take into account the effect of dissipation and temperature at the same time. 
Non-equilibrium phase transitions and critical phenomena have been intensively studied in driven dissipative classical many-body systems \cite{denBroeckC94, ChakrabartiB99, FujisakaH01, DuttaS04, PopkovV08} and our theory can provide a reliable framework for predicting and even controlling them.

Second, the master equation of a classical stochastic system is a prototypical example of a non-Hermitian Schr$\mr{\ddot{o}}$dinger equation as mentioned in Sec.~\ref{sec: FME, cla}. 
Recently considerable efforts have been devoted to exploring non-Hermitian physics both experimentally and theoretically, in particular their topological aspects \cite{SchomerusH13, MalzardS15, LeeT16, MuruganA17, LeykamD17, XuY17, ZhouH18, BandresM18, DasbiswasK18}. 
Remarkably, topological classifications of static non-Hermitian systems \cite{GongZ18, KawabataK18} has been found to be significantly different from the Hermitian counterpart \cite{RyuS10, SchnyderA09, KitaevA09}. 
Since Hermitian Floquet systems exhibit unique topological phenomena which have no counterparts in static ones, e.g., anomalous edges \cite{KitagawaT10b, JiangL11, RudnerM13} and gapless lattice-prohibited bands \cite{Thouless83a, KitagawaT10b, BudichJ17, HigashikawaS18}, it is natural to expect that non-Hermitian Floquet systems also host unique topological phases, which are different from both Hermitian Floquet systems and non-Hermitian static ones.  
While non-Hermitian Floquet systems have been studied in the context of quantum walks \cite{RudnerM09, KimD16, XiaoL17}, their realizations and properties of topological edge states in classical stochastic systems are largely unexplored. 

\section{acknowledgement}
We acknowledge Zongping Gong, Kazuya Fujimoto, Ryusuke Hamazaki, Fumihiro Ishikawa, Tatsuhiko N. Ikeda, Takashi Mori, Tatsuhiko Shirai, Masaru Hongo, Sota Kitamura, Takashi Oka, Alexander Schnell, and André Eckardt for fruitful discussions. 
S. H and H. F. are supported by Advanced Leading Graduate Course for Photon Science (ALPS) of Japan Society for the Promotion of Science (JSPS). S.H is supported by JSPS KAKENHI Grant-in-Aid for JSPS Fellows Grant No.~JP16J03619. H.F is supported by JSPS KAKENHI Grant-in-Aid for JSPS Fellows Grant No.~JP16J04752. M. S. was supported by Grant-in-Aid for Scientific Research on Innovative Area, ”Nano Spin Conversion Science” (Grant No.17H05174), and JSPS KAKENHI (Grant No. JP17K05513 and No. JP15H02117). A part of the computation in this work has been done using the facilities of the Supercomputer Center, the Institute for Solid State Physics, the University of Tokyo.

\appendix
\section{\label{app: der, sLLG}FM expansion of the sLLG equation}
In this Appendix, we explain how to compute the low-order terms of the FM expansion for the sLLG equations 
describing laser-driven magnets in Secs.~\ref{sec: sLLG} and \ref{sec: spintronics}.

\subsection{\label{app: der, sLLG, 1stFM}Two-dimensional ferromagnet: first order}
We first consider the first-order term of the FM expansion for the sLLG equation 
in Sec.~\ref{sec: sLLG}. For a systematic calculation of commutators in the FM expansion, 
we rewrite the Fokker-Planck operator in terms of the angular momentum operators. 
Let us define the angular momentum operators $L_{\bmr, a}$ and the operators $K_{\bmr, a}$ and $N_{\bmr, a}$ by 
\begin{align}
	L_{\bmr, a} &:= - \epsilon_{abc} 
	m_{\bmr, b} {\partial \over \partial m_{\bmr, c}}
	, 
	\nonumber \\
	K_{\bmr, a} &:= \epsilon_{abc} L_{\bmr, b} m_{\bmr, c}
	, \nonumber \\
	N_{\bmr, a} &:= L_{\bmr, a} + \alpha K_{\bmr, a}
	, 
\end{align} 
where $\epsilon_{abc}$ is the totally antisymmetric tensor of rank $3$, and $\alpha$ is the Gilbert damping in Eq.~\eqref{eq: sLLG, original}. 
These operators satisfy the following commutation relations: 
\begin{align}
	\rbr{L_{\bmr, a}, L_{\bmr\pri, b}} &= 
	\delta_{\bmr, \bmr\pri} \epsilon_{abc} L_{\bmr, c}
	, \nonumber \\
	\rbr{K_{\bmr, a}, K_{\bmr\pri, b}} &=
	- \delta_{\bmr, \bmr\pri} \epsilon_{abc} K_{\bmr, c}
	, \nonumber \\
	\rbr{L_{\bmr, a}, K_{\bmr\pri, b}} &=
	\delta_{\bmr, \bmr\pri} \epsilon_{abc} L_{\bmr, c}
	, \nonumber \\
	\rbr{N_{\bmr, a}, N_{\bmr\pri, b}} &= 
	\delta_{\bmr, \bmr\pri} \tbr{
	\epsilon_{abc} N_{\bmr, c} + 
	\alpha \br{
	N_{\bmr, a} m_{\bmr, b} - m_{\bmr, a} N_{\bmr, b}
	}
	}
	, \label{eq: app, B, comm}
\end{align}
Consider the commutator between the operators $\mL_\alpha := - \sum_{\bmr} \mr{div}[ \bmf_{\bmr, \alpha} \quad \cdot \quad]$ and $\mL_\beta := - \sum_{\bmr} \mr{div}[ \bmf_{\bmr, \beta} \quad \cdot \quad]$, 
where $\bmf_{\bmr, \gamma}$ ($\gamma = A, B$) is given by 
\begin{align}
	\bmf_{\bmr, \gamma} &=
	 - {\bmm_{\bmr} \over 1 + \alpha^2} 
	 \times 
	\br{ \bm{H}_{\bmr, \gamma} + {\alpha \over m_s} \bmm_{\bmr} \times \bm{H}_{\bmr, \gamma}}
	. 
\end{align}
From the equation
\begin{align}
	{\partial \over \partial m_a} \br{
	\epsilon_{abc} m_{\bmr, b} \bm{H}_{\bmr, c}
	} &= 
	- \epsilon_{abc}
	m_{\bmr, a}
	{\partial \over \partial m_{\bmr, b}} \br{ 
	\bm{H}_{\bmr, \gamma, c}
	}
	\nonumber \\
	 & = \bm{L}_{\bmr}\cdot \bm{H}_{\bmr, \gamma}
	, 
\end{align}
we can rewrite $\mL_\gamma$ in terms of $\bm{N}_{\bmr}$ as follows: 
\begin{align}
	\mL_{\gamma} &= 
	\sum_{\bmr}
	\br{\bm{L}_{\bmr} \cdot \bar{\bm{H}}_{\bmr, \gamma}
	 + \alpha \
	\bm{K}_{\bmr} \cdot \bar{\bm{H}}_{\bmr, \gamma}}
	 = \sum_{\bmr}
	 \bm{N}_{\bmr} \cdot \bar{\bm{H}}_{\bmr, \gamma}
	, 
\end{align}
where $\bar{\bm{H}}_{\bmr, \gamma} = \bm{H}_{\bmr, \gamma} / (1 + \alpha^2)$. 
Using Eq.~\eqref{eq: app, B, comm}, 
we obtain 
\begin{align}
	&\rbr{ \mL_A, \mL_B}
	 = \sum_{\bmr} \bm{N}_{\bmr} 
	 \rbr{ \bar{\bm{H}}_A, \bar{\bm{H}}_B}_{\bmr, \mr{mag}}
	 ,
\end{align}
where commutator $\rbr{ \bar{\bm{H}}_A, \bar{\bm{H}}_B}_{\bmr, \mr{mag}}$ is defined by 
\begin{align}
	 & \rbr{ \bar{\bm{H}}_A, \bar{\bm{H}}_B}_{\bmr, \mr{mag}}
	 \nonumber \\
	 :=& 
	 \bar{\bm{H}}_{\bmr, A} \times \bar{\bm{H}}_{\bmr, B}
	  + {\alpha \bmm_{\bmr} \over m_s} \times 
	 \br{
	 \bar{\bm{H}}_{\bmr, A} \times \bar{\bm{H}}_{\bmr, B}
	 }
	 \nonumber \\
	  & + \sum_{\bmr\pri} \rbr{\br{
	  \bar{\bm{H}}_{\bmr\pri, A} \cdot \bm{L}_{\bmr\pri}
	  }\bar{\bm{H}}_{\bmr, B} - 
	  \br{
	  \bar{\bm{H}}_{\bmr\pri, B} \cdot \bm{L}_{\bmr\pri}
	  }\bar{\bm{H}}_{\bmr, A} 
	  }
	 \nonumber \\
	  & + {\alpha \over m_s}
	  \sum_{\bmr\pri} \left[\br{
	  \bmm_{\bmr\pri} \cdot
	  \bar{\bm{H}}_{\bmr\pri, A} \times \bm{L}_{\bmr\pri}
	  }\bar{\bm{H}}_{\bmr, B} - 
	  \right.
	  \nonumber \\ 
	  &\quad \quad \quad \quad \quad 
	  \left. 
	  \br{
	  \bmm_{\bmr\pri} \cdot
	  \bar{\bm{H}}_{\bmr\pri, B} \times \bm{L}_{\bmr\pri}
	  }\bar{\bm{H}}_{\bmr, A} 
	  \right]
	  . \label{eq: magnetic, commutator}
\end{align}
Thus, $\rbr{ \mL_A, \mL_B}$ defines the drift field with magnetic field $\rbr{ \bar{\bm{H}}_A, \bar{\bm{H}}_B}_{\bmr, \mr{mag}}$. 
For example, when $\bm{H}_{\bmr, A} = \bm{B}_{-1}$ and $\bm{H}_{\bmr, B} = \bm{B}_{+1}$, 
we obtain 
\begin{align}
	\rbr{ \bar{\bm{H}}_A, \bar{\bm{H}}_B}_{\bmr, \mr{mag}}
	 &= { \bm{B}_{-1} \times \bm{B}_{+1} \over (1 + \alpha^2)\omega} + {\alpha \bmm_{\bmr} \over m_s } \times 
	 { \bm{B}_{-1} \times \bm{B}_{+1} \over (1 + \alpha^2)\omega} 
	 \nonumber \\
	 & = \bm{b}^{(1)} + {\alpha \bmm_{\bmr} \over m_s } 
	  \times \bm{b}^{(1)}
	. 
\end{align}
Therefore, from the first-order FM expansion, 
we obtain 
\begin{align}
	\mL_F^{(1)} &:= 
	\sum_{\bmr}
	\bm{L}_{\bmr} \cdot 
	\br{
	\bm{b}^{(1)} + {\alpha \bmm_{\bmr} \over m_s } 
	  \times \bm{b}^{(1)}
	  }
	, 
\end{align}
which gives Eq.~\eqref{eq: sLLG, FM, 1stfield}.

\subsection{Two-dimensional ferromagnet: second order}
The second-order FM expansion $\mL_F^{(2)}$ is given by 
\begin{align}
	\mL_F^{(2)} &= 
	- {
	\rbr{\mL_{-1}\rbr{\mL_0, \mL_1}} + 
	\rbr{\mL_{1}\rbr{\mL_0, \mL_{-1}}}
	 \over 2\omega^2}
	. 
\end{align}
We decompose $\mL_0$ into the terms on external field, the nearest-neighbor interaction, and diffusion: 
\begin{align}
	\mL_0^{\mr{ext}} &= {1 \over 1 + \alpha^2} 
	\sum_{\bmr} \bm{L}_{\bmr} \cdot \bm{B}_0
	, \nonumber \\
	\mL_0^{\mr{int}} &= {J \over 1 + \alpha^2} \sum_{\bmr} \bm{L}_{\bmr} \cdot \sum_{\bmr\pri: n.n} \bmm_{\bmr\pri}
	, \nonumber \\
	\mL_0^{\mr{dif}} &= \mr{div}_2\br{\mathcal{D} \ \ \cdot \ \ }
	, \label{eq: app, mL0}
\end{align}
and decompose $\mL_F^{(2)}$ accordingly: $\mL_F^{(2)} = \mL_F^{(2), \mr{ext}} + \mL_F^{(2), \mr{int}} + \mL_F^{(2), \mr{dif}}$. 
Combining Eqs.~\eqref{eq: magnetic, commutator} and \eqref{eq: app, mL0}, we obtain 
\begin{align}
	\mL_F^{(2), \mr{ext}} &= 
	\sum_{\bmr}
	\bm{L}_{\bmr} \cdot 
	\br{(1-\alpha^2)
	\bm{b}^{(2)} + {2 \alpha \bmm_{\bmr} \over m_s } 
	  \times \bm{b}^{(2)}
	  }
	, \nonumber \\
	\mL_F^{(2), \mr{int}} &= 
	\sum_{\bmr}
	\bm{L}_{\bmr} \cdot 
	J \br{{\alpha B_d \over m_s \omega (1 + \alpha^2)}}^2
	m_{\bmr\pri, z}
	\nonumber \\
	& \quad \times
	\sum_{\bmr\pri: n,n}
	\thvec{m_{\bmr\pri, x} \delta m_{\bmr, \bmr\pri,z}}
	{m_{\bmr\pri, y} \delta m_{\bmr, \bmr\pri,z}}
	{- m_{\bmr\pri, z} \delta m_{\bmr, \bmr\pri,x} - 
	m_{\bmr\pri, y} \delta m_{\bmr, \bmr\pri,y}}	
	\nonumber \\
	 &= \sum_{\bmr}
	\bm{L}_{\bmr} \cdot 
	\sum_{\bmr\pri: n,n} \bm{\delta} \bm{J}_{\bmr, \bmr\pri}
	, \nonumber \\
	\mL_F^{(2), \mr{dif}} &= 
	\mr{div}\rbr{
	\br{
	{2 \chi_{\bmr} \over 1 + \alpha^2} \bmm_{\bmr}} \ \cdot \ 
	} + \mr{div}_2 \rbr{
	\chi_{\bmr} D G G^{\mr{tr}}  \ \cdot \
	}
	, 
\end{align}
where $\delta \bmm_{\bmr, \bmr\pri} = \bmm_{\bmr} - \bmm_{\bmr\pri}$. 
The overall effective master equation is given by 
\begin{align}
	\partial_t P &= 
	\sum_{\bmr}\br{
	\bm{L}_{\bmr} \cdot \bm{H}\pri_{F, \bmr} P
	} + 
	\mr{div}\rbr{
	\br{
	{2 (1+\chi) \over 1 + \alpha^2} \bmm_{\bmr}} P
	} 
	\nonumber \\
	&\quad + \mr{div}_2 \rbr{
	(1 + \chi) D G G^{\mr{tr}} P
	}
	, \label{eq: FP, sLLG, 2nd} 
\end{align}
where the effective field $\bm{H}\pri_{F, \bmr}$is 
\begin{align}
	\bm{H}\pri_{F, \bmr}  
	 &= 
	\sum_{\bmr\pri: n,n} \br{
	J \bmm_{\bmr} + \bm{\delta} \bm{J}_{\bmr, \bmr\pri}
	} + 
	\bm{B}_0 + \bm{b}^{(1)} + (1 - \alpha^2) \bm{b}^{(2)}
	 \nonumber \\
	&\quad + \br{
	 - {\alpha \over m_s} \bm{b}^{(1)} 
	 - {2 \alpha \over m_s} \bm{b}^{(2)}
	} \times \bmm_{\bmr}
	. \nonumber \\
\end{align}
By defining a new diffusion matrix $G_F$ by $G_F := (1 + \chi )^{1/2} G$, 
the second and third terms on the right-hand side of Eq.~\eqref{eq: FP, sLLG, 2nd} are rewritten as 
\begin{align}
	&\mr{div}\rbr{
	\br{
	{2 (1+\chi) \over 1 + \alpha^2} \bmm_{\bmr}} P
	}
	\nonumber \\
	& = 
	\sum_{\bmr}
	\bm{L}_{\bmr} \cdot 
	\br{
	- \bm{b}^{(2), \mr{dif}}
	+ {\bm{b}^{(2), \mr{dif}} \over m_s} \times \bmm_{\bmr}
	} P \nonumber \\
	& \quad + 
	\mr{div}
	\br{
	- \bm{d}_F P
	} 	
	, \nonumber \\
	&\mr{div}_2 \rbr{
	(1 + \chi) D G G^{\mr{tr}} P
	} = \mr{div}_2 \rbr{
	D G_F G_F^{\mr{tr}} P
	}
	, 
\end{align}
where $\bm{d}_{F,i} := g_{F, kl} \partial_k g_{F, il}$ and 
\begin{align}
	\bm{b}^{(2), \mr{dif}} := 
	{ D m_s \over 2(1+ \alpha^2 )} {\delta \chi_{\bmr} \over \delta \bmm_{\bmr}}
	. 
\end{align}
Equation \eqref{eq: FP, sLLG, 2nd} is then rewritten as 
\begin{align}
	\partial_t P &= 
	\sum_{\bmr}\br{
	\bm{L}_{\bmr} \cdot \bm{H}_{F, \bmr} P
	} + 
	\mr{div}
	\br{
	- \bm{d}_F P
	} 
	\nonumber \\
	&\quad + \mr{div}_2 \rbr{
	D G_F G_F^{\mr{tr}} P
	}
	, 
\end{align}
where $\bm{H}_{F, \bmr}$ is the effective field defined in 
Eq.~\eqref{eq: effective field FM2}. 
Comparing this equation with Eq.~\eqref{eq: fokker, coef}, 
we finally arrive at the sLLG equation \eqref{eq: eff, sLLG, 2nd}.

\subsection{Multiferroic spin chain}
The FM expansion of the sLLG equation for the multiferroic spin chain model~(\ref{eq:MultiferroModel}) can be performed in a manner similar to what we have done in  App.~\ref{app: der, sLLG, 1stFM}. 
The first-order Fokker-Planck operator $\mL_F^{(1)}$ is given by 
\begin{align}
	\mL_F^{(1)} &:= 
	\sum_{\bmr}
	\bm{L}_{\bmr} \cdot 
	{i \over \omega}
	\rbr{ \bar{\bm{H}}_{-1}, \bar{\bm{H}}_1}_{\bmr, \mr{mag}}
	, 
\end{align}
where the Fourier harmonics $\bar{\bm{H}}_{\pm 1}$ of the effective field is 
\begin{align}
	\bar{\bm{H}}_{\pm 1} &:= 
	\sum_{\bmr\pri: n.n.} \bm{D}_{\pm} \times \bmm_{\bmr\pri} + \bm{B}_{\pm}
	, \nonumber \\
	\bm{D}_{\pm} &:= 
	{g_{me} E_d \over 2} \br{\pm i, 1, 0}^{\mr{tr}}
	, \nonumber \\
	\bm{B}_{\pm} &:= 
	{B_d \over 2} \br{1, \mp i, 0}^{\mr{tr}}
	. 
\end{align}
By a straightforward calculation, we obtain 
\begin{align}
	\mL_F^{(1)} &:= 
	\sum_{\bmr}
	\bm{L}_{\bmr} \cdot \bm{H}_{F, \bmr}^{(1)}
	, \nonumber \\
	\bm{H}_{F, \bmr}^{(1)} &:= 
	\sum_{\bmr\pri: n.n.} 
	\br{
	\bm{D}_{F, \bmr, \bmr\pri} \times \bmm_{\bmr\pri}
	} + \bm{B}_F - \alpha \bm{B}_F \times \bmm_{\bmr}
	 \nonumber \\
	& - \sum_{\bmr\pri: n.n.} 
	{\alpha \epsilon_B\epsilon_E \over 2 m_s (1+  \alpha^2) \omega}
	  \thvec{ m_s^2 + m_{\bmr\pri, y}\delta m_{\bmr, y}}{- m_{\bmr\pri, y}\delta m_{\bmr, x}}{0}
	, 
\end{align}
where $\bm{D}_{F, \bmr, \bmr\pri}$ and $\bm{B}_F$ are defined in Eq.~\eqref{eq: mf, eff_field, FM}. 
This equation follows from the correspondence between the master equation and the EOM.


\bibliography{DD_higashi}

\end{document}